\documentclass[twocolumn]{article}
\usepackage{graphicx} % Required for inserting images 
\usepackage{todonotes}
\usepackage{subcaption}
\usepackage{caption}
\usepackage[margin=2.5cm]{geometry}
\usepackage{amssymb}
\usepackage{amsmath}
\usepackage{bbm} % for \mathbbm{1} indicator function
\usepackage{tikz}
\usetikzlibrary{arrows.meta,positioning,shapes.geometric}
\usepackage{listings}
\usepackage{multirow}
\usepackage{booktabs}
\usepackage{tabularx} % automatic line-wrapping columns in tables
\usepackage{pdflscape} % landscape pages for wide tables/figures
\usepackage{subcaption}
\usepackage{float}
\usepackage{dblfloatfix} % improves placement of two-column floats
\usepackage[section]{placeins} % keep floats within their section
\usepackage{enumitem}
\usepackage{authblk}

\usepackage[colorlinks=true,linkcolor=blue,citecolor=blue,urlcolor=blue]{hyperref}
\usepackage[noabbrev,nameinlink,capitalise]{cleveref}

\title{Evaluating Generalization Mechanisms in Autonomous Cyber Attack Agents}
\author[1]{Luk{\'a}{\v{s}} Ond{\v{r}}ej}
\author[2]{Jihoon Shin}
\author[2]{Emilia Rivas}
\author[3]{Diego Forni}
\author[1]{Maria Rigaki}
\author[3]{Carlos Catania}
\author[2]{Aritran Piplai}
\author[2]{Christopher Kiekintveld}
\author[1]{Sebastian Garc{\'i}a}

\affil[1]{Czech Technical University in Prague (CTU)}
\affil[3]{Universidad Nacional de Cuyo (UNCUYO)}
\affil[2]{University of Texas at El Paso (UTEP)}

\date{March 2026}

\begin{document}

\maketitle
\begin{abstract}
Autonomous offensive agents often fail to transfer beyond the networks on which they are trained. We isolate a minimal but fundamental shift---unseen host/subnet IP reassignment in an otherwise fixed enterprise scenario---and evaluate attacker generalization in the NetSecGame environment. Agents are trained on five IP-range variants and tested on a sixth unseen variant; only the meta-learning agent may adapt at test time. We compare three agent families (traditional RL, adaptation agents, and LLM-based agents) and use action-distribution-based behavioral/XAI analyses to localize failure modes. Some adaptation methods show partial transfer but significant degradation under unseen reassignment, indicating that even address-space changes can break long-horizon attack policies. Under our evaluation protocol and agent-specific assumptions (Table~\ref{tab:comparability}), prompt-driven pretrained LLM agents achieve the highest success on the held-out reassignment, but at the cost of increased inference-time compute, reduced transparency, and practical failure modes such as repetition/invalid-action loops.
\end{abstract}

% ---------------------------------------
\section{Introduction}
% What is the Problem
Autonomous offensive agents trained in simulated cybersecurity network games often learn policies that are brittle when the network changes at test time \cite{Wolk2022,Janisch2023a}. Even when the attacker objective is unchanged, small shifts in network identifiers can invalidate a learned sequence of reconnaissance, exploitation, and lateral movement actions \cite{becker2024evaluation,microsoft2021cyberbattlesim}. This limits the practical value of cyber agents, since enterprise networks routinely apply address reassignment, segmentation, and reconfiguration.

\paragraph{Problem}
We study whether attacker policies can generalize under a minimal, controlled form of shift: unseen IP address space changes through host and subnet IP reassignment, applied to an otherwise fixed enterprise scenario. In this paper, \emph{generalization} means training on several IP-range variants of the same scenario and then performing well under a previously unseen reassignment. We measure the seen-to-unseen drop using win rate, return, and steps.

\paragraph{Limitations of prior approaches}
Prior work on autonomous cyber agents often evaluates on fixed networks, where policies can inadvertently memorize concrete identifiers \cite{Applebaum2022,Wolk2022}. Even when neural representations are used, identifier tokens can dominate similarity structure, so semantically equivalent states become far apart under reassignment \cite{Nyberg2024}. Abstract policies based on graph and entity representations \cite{Collyer2022,Symes2024,King2025} and fast adaptation methods such as meta-learning \cite{finn2017model} aim to mitigate this, but results are hard to interpret when multiple sources of shift change at once.

\paragraph{Proposal, hypotheses, and method}
In this work, we analyze attacker generalization in NetSecGame~\cite{NetSecGame_code} under controlled address-space shift. We generate multiple variants of the same enterprise scenario that differ only by host and subnet IP reassignment, train agents on five variants, and evaluate on a sixth unseen variant. We compare six attackers grouped into three families: (i) \textbf{traditional RL} (DQN-style value learning and a DDQN+embedding variant), (ii) \textbf{LLM-based} decision making built on ReAct-style prompting \cite{happe2023getting,yao_react_2022}, and (iii) \textbf{generalization and adaptation} methods based on conceptual abstraction and gradient-based meta-learning \cite{finn2017model}.

Our hypotheses are:
\begin{itemize}
    \item \textbf{H1 Identifier-dependent policies fail:} Agents whose internal state and action representations depend on concrete IP and subnet values will exhibit a large performance drop from seen networks to the unseen IP reassignment.
    \item \textbf{H2 Abstraction and adaptation reduce the drop:} Address-invariant abstractions and test-time adaptation will reduce the seen-to-unseen drop relative to traditional RL baselines.
\end{itemize}

Agents are trained on five IP-range variants and evaluated on a sixth unseen variant. Inputs are limited to what is discoverable through interaction. During evaluation, only meta-learning agents are allowed test-time parameter updates; all other agents act as frozen policies.

\paragraph{Results and analysis summary}
We find that address-space reassignment is sufficient to break strong seen-topology performance for identifier-dependent agents, supporting H1. In contrast, conceptual abstraction transfers substantially better, and meta-learning enables partial recovery via test-time adaptation, supporting H2. LLM-based agents can achieve high success on the unseen topology but exhibit distinctive failure modes such as invalid-action loops, motivating behavioral analyses beyond aggregate win rate.

Even minimal IP reassignment can induce a large generalization gap in long-horizon cyber attack policies. Robust transfer appears to require representations and decision mechanisms that are explicitly address-invariant and/or capable of efficient test-time adaptation.

%%%%%%%%%%%%%%%%%%%%%%%%%%%%%%%%%%%%%%%
\section{Motivation and Background}

IP reassignment induces a minimal but fundamental form of distribution shift for cyber agents. The enterprise scenario can remain logically unchanged, but the attacker observes and acts through concrete identifiers that have been permuted. As a result, a policy that binds decisions to specific IP values may fail even when the same functional attack strategy should apply.

A useful way to interpret this failure mode is through analogies from embodied reinforcement learning. DARLA~\cite{higgins2017darla} highlights how policies can overfit to surface cues in one visual domain and then break under appearance shift. The cyber analogue is overfitting to identifiers: an attacker may implicitly learn a brittle routine such as pivoting through a particular address, instead of targeting hosts based on their functional role, reachability, and discovered attack surface.

This shift is non-trivial under partial observability. In NetSecGame, hosts, services, and connectivity are discovered through interaction, and actions are parameterized by the set of identifiers observed so far. Generalization under IP reassignment therefore requires learning address-invariant role information from incomplete evidence, and selecting valid actions in a changing candidate set.

\begin{figure}[t]
    \centering
    \includegraphics[width=0.75\linewidth]{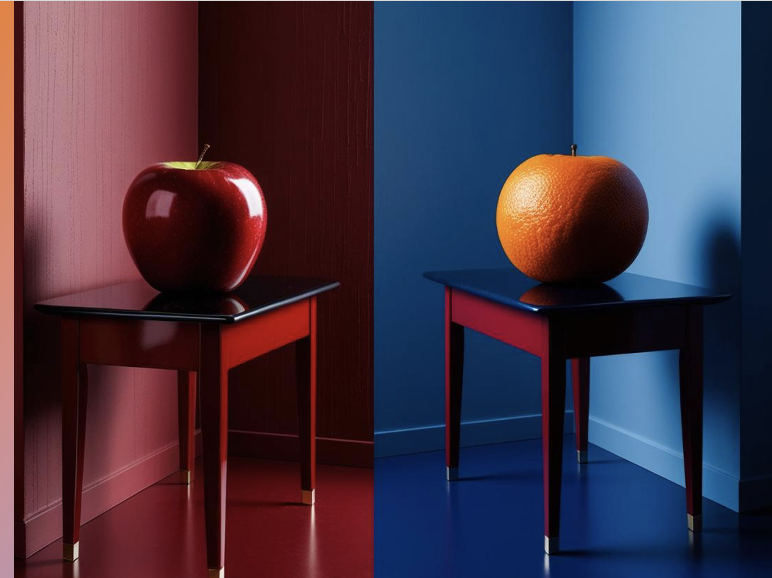}
    \caption{\textbf{DARLA \cite{higgins2017darla} intuition for generalization:} A robot trained in one visual domain (e.g., a red room with a red fruit) can fail under appearance shifts (e.g., a blue room with an orange fruit) if it overfits to surface cues.}
    \label{fig:darla}
\end{figure}

\subsection{Address-invariant representations}
A complementary approach is to reduce dependence on concrete identifiers by changing what the agent represents and reasons over. Graph- and entity-based policies aim to encode network structure in a way that is less sensitive to renaming hosts or subnets \cite{Collyer2022,Nyberg2024,Symes2024}. In our setting, the goal of abstraction is to preserve the information needed for attack planning, such as reachability, privilege state, and service exposure, while discarding the specific IP and subnet values that change under reassignment.

\subsection{Test-time adaptation via meta-learning}
One approach to handle shift is to allow limited learning in the new environment. Gradient-based meta-learning methods such as Model-Agnostic Meta Learning (MAML) learn an initialization that can adapt to a new task in a few gradient steps using a small support set \cite{finn2017model}. Reptile is a first-order alternative that also learns a transferable initialization but avoids second-order derivatives \cite{nichol2018first}. In this paper, meta-learning is used as an explicit adaptation mechanism under unseen IP reassignment: unlike other agents, the meta-learning agents are allowed to update their parameters at test time before evaluation.

These perspectives motivate our comparative analysis of how traditional RL, abstraction mechanisms, meta-learning, and LLM-based decision making behave under controlled IP reassignment. We next summarize prior work on topology variation, realistic cyber environments, and representation choices for transfer.

\section{Previous Work}
\label{sec:prev_work}

\subsection{Changing Network Topologies} 
Early cyber RL agents were evaluated on fixed topologies, limiting their ability to generalize. More recent work varies in the network structure or size to test transfer. Collyer et al.~\cite{Collyer2022} randomized connectivity while keeping node counts fixed, while Applebaum et al.~\cite{Applebaum2022} and Mern et al.~\cite{Mern2021} varied the number of hosts. The CAGE-4 competition explicitly introduced variable user machines, and Wolk et al.~\cite{Wolk2022} showed significant performance drops when defenders faced unseen topologies. These results highlight the brittleness of fixed-topology policies.

\subsection{Goals and Realism} 
Static goals lead to overfitting. CyberBattleSim randomizes target placement~\cite{King2025}, while Janisch et al.~\cite{Janisch2023a} enable training in simulation and deployment in emulation with altered networks and goals. Modern environments also emphasize realism: CybORG simulates enterprise domains, Mern et al.~\cite{Mern2022} use ICS testbeds, and Janisch et al.~\cite{Janisch2023a} integrate VirtualBox and Metasploit. This reflects a move from toy abstractions toward realistic, variable settings.

\subsection{Abstract Policy Representations} 
Generalization requires abstract policies. Graph neural networks (GNNs) are widely applied: Collyer et al.~\cite{Collyer2022} embed network graphs, Janisch et al.~\cite{Janisch2023b} use relational GNNs with autoregressive policies, and Nyberg \& Johnson~\cite{Nyberg2024} apply message passing. Entity-based policies are also effective: Symes Thompson et al.~\cite{Symes2024} use a Transformer over network entities, achieving permutation invariance. Hierarchical and ensemble methods appear in CAGE~\cite{Wolk2022} to combine specialized sub-policies.

Table~\ref{tab:prior_work} provides a compact summary of representative recent papers on autonomous cyber agents, organized by environment, training regime, and whether evaluation includes tests on unseen network variants. The goal of this table is not to exhaustively cover the literature, but to make explicit which dimensions of generalization are exercised in prior evaluations. In particular, many studies vary connectivity or host counts and report aggregate performance, but differ in whether they report numeric transfer gaps and whether the unseen tests isolate a single source of shift. This motivates our controlled IP reassignment setting, where only addressing changes while the enterprise scenario remains fixed.
\begin{table*}[ht]
\centering
\scriptsize
\setlength{\tabcolsep}{3pt}
\caption{Summary of environments, training regimes (TR), unseen topology tests (UT), performance, and numerical results (Num. res.) in recent work on autonomous cybersecurity agents. TR: NT/ep.=new topology each episode, FT$\rightarrow$NT=fixed then new, Size scale=trained on smaller networks then larger, Multi-ep./topo=many episodes per topology, Fixed=no change. UT: \checkmark=yes, $\times$=no, (\checkmark)=partial. Num. res.: \checkmark=numeric results reported, $\times$=no. n/r=not reported, var.=variable.}
\begin{tabularx}{\textwidth}{l|X|c|c|c|X|c}
\textbf{Paper} & \textbf{Environment} & \textbf{Hosts} & \textbf{TR} & \textbf{UT} & \textbf{Performance} & \textbf{Num. res.} \\
\hline
Entity \cite{Symes2024} & Yawning Titan sim. & n/r & NT/ep. & \checkmark & Entity Transformer $>$ MLP baseline & $\times$ \\
Graph \cite{King2025} & Graph-based ACD & 3--10 & FT$\rightarrow$NT & \checkmark & $\sim$4$\times$ baseline score & \checkmark \\
Relat. \cite{Janisch2023b} & Relational SysAdmin-style env. & var. & Size scale & \checkmark & Zero-shot transfer reported & $\times$ \\
ACD-G \cite{Collyer2022} & ACD-G custom env. & 8--16 & Multi-ep./topo & \checkmark & Generalization improves over baseline & $\times$ \\
ICS \cite{Mern2022} & ICS testbed & $\sim$10 & Fixed & (\checkmark) & Autonomous mitigation demonstrated & \checkmark \\
Attack \cite{Oh2023} & Simulated attack env. & n/r & Fixed & $\times$ & 78\% success rate & \checkmark \\
TabQ \cite{Applebaum2022} & CyberBattleSim variants & 4--17 & Multi-ep./topo & \checkmark & Abstractions outperform targeted policies on unseen & \checkmark \\
NASimEmu \cite{Janisch2023a} & NASimEmu sim+emu & var. & NT/ep. & \checkmark & Generalizes to novel scenarios; runs in emulation & \checkmark \\
ICS-Orch \cite{Mern2021} & ICS simulation & $\sim$83 & Fixed & $\times$ & Attention DQN $>$ conv baseline; faster convergence & \checkmark \\
CAGE \cite{Wolk2022} & CybORG CAGE-2 & 13 & Multi-ep./topo & \checkmark & Ensembles strong but drop on unseen networks/attacks & \checkmark \\
MPNN \cite{Nyberg2024} & CybORG CAGE-2 variants & 13 & NT/ep. & \checkmark & Message-passing transfers; MLP needs retraining & \checkmark \\
\end{tabularx}
\label{tab:prior_work}
\end{table*}

Table~\ref{tab:prior_work} now covers every paper that is discussed in Section~\ref{sec:prev_work}, so each cited model is paired with its environment, regime, generalization test, and reported numerical results for easy comparison.

\subsection{Value-based Reinforcement Learning Attackers}
Deep Q-Learning (DQN) has been applied across various domains, including video games \cite{mnih2013playing}, computer vision \cite{chipka2021computer}, and healthcare \cite{abdellatif2023reinforcement}. In cybersecurity, value-based RL has been used for sequential attack planning in environments such as NASim and CyberBattleSim. Comparative studies \cite{becker2024evaluation} benchmark DQN against other RL methods on NASim tasks, and two-player formulations extend DQN to attacker--defender games in CyberBattleSim. Piplai et al.~\cite{piplai2022knowledge} introduce a knowledge-guided two-player minimax DQN built on CyberBattleSim; while useful for benchmarking, such environments may abstract away network structure (e.g., segmentation) in ways that can simplify exploration \cite{microsoft2021cyberbattlesim}. These lines of work motivate our inclusion of DQN-style agents as representative traditional methods for long-horizon cyber attack tasks.

\subsection{Reinforcement Learning Transfer to Unseen Tasks}
RL transfer to unseen tasks aims to train agents that can solve new tasks or environments without additional online interaction, often by learning task-agnostic representations during a reward-free pre-training phase ~\cite{kirk2023survey}. Recent work on transfer from low-quality offline data uses conservative value regularization to stabilize zero-shot performance across downstream tasks defined over a fixed dataset of environment interactions ~\cite{jeen2024zero}. Extensions of forward–backward models to partial observability augment these representations with recurrent memory, enabling agents to better handle noisy, corrupted, or intermittently missing observations during zero-shot deployment ~\cite{jeen2025zero}. Other approaches emphasize architectural inductive biases for transfer, such as attention-based perception for visual zero-shot control or graph-based policies that generalize to previously unseen topologies ~\cite{genc2020zero,chen2021zero}. Sim-to-real transfer methods like Siamese-Q further demonstrate that policies trained entirely in simulation can be deployed zero-shot to real systems via representation learning that aligns simulated and real observations ~\cite{zhang2025zero,kirk2023survey}.

\subsection{Meta Learning in Cyber Battle Simulation}
Meta-learning has begun to be applied as a foundational approach to overcome the challenges in cyberspace security, particularly addressing the challenges of adapting to evolving attack modes where limited training samples are available \cite{yang2023application}. The Generalizable Autonomous Penetration (GAP) framework applies Meta-RL with the MAML algorithm to enable zero-shot policy transfer and rapid adaptation in autonomous penetration testing, providing a pathway to robust and realistic cyber simulations \cite{zhou2024mind}. In addition, the Robust Meta Network Embedding (ROMNE) framework integrates adversarial attack and defense models into a dynamic adversarial training process, improving the robustness of network embedding against constant noise in adversarial environments \cite{zhou2020robust}. These works demonstrate the potential of meta-learning in advancing both offensive and defensive strategies in cyber battle simulations. They provide the foundation for developing MAML-based attacker agents explored in this study.

\subsection{Limitations and positioning}
Prior work has made progress by varying connectivity and host counts \cite{Collyer2022,Applebaum2022,Mern2021,Wolk2022}, by increasing realism through simulation-to-emulation or richer enterprise settings \cite{Janisch2023a,Mern2022}, and by introducing abstract policy representations such as graph and entity encodings \cite{Janisch2023b,Nyberg2024,Symes2024}. However, generalization results are often difficult to interpret because multiple sources of shift change simultaneously, for example topology size, target placement, observation encodings, or objectives.

Our work complements these directions by isolating a single, practically relevant source of shift: host and subnet IP reassignment in an otherwise fixed enterprise scenario. This controlled setting enables a direct comparison of generalization and adaptation mechanisms across diverse attacker designs under the same threat model and partial-observability constraints, and supports diagnosing failures using behavioral analyses rather than only aggregate success metrics.

\section{NetSecGame Environment}
The NetSecGame environment (NSG)~\cite{NetSecGame_code} is a free-software, multi-agent cybersecurity simulation environment built with parts of the CYST network simulator~\cite{cyst_core}. It provides an interface for attacker and defender agents to interact with a configurable enterprise network, where actions correspond to common cyber operations such as scanning, service discovery, exploitation, and data exfiltration. Scenarios are specified through configuration files (e.g., network hosts and services, firewall rules, goals, and reward shaping), which makes it possible to generate controlled task variants---including the host/subnet IP reassignment used in this paper---while keeping the underlying enterprise structure fixed.

At each step, NSG exposes to an agent only the information it has discovered so far (e.g. known hosts), and it accepts parameterized actions whose arguments must be drawn from that current knowledge. This observation/action design matches the sequential structure of real attacks and makes address-level shifts directly affect which state and action signatures an agent encounters.

\subsection{State and Action Space}
Each state observed by an agent on each step in NSG is represented by the set of assets it has discovered so far. These assets determine which parameterized actions are available:
\begin{enumerate}
    \item \textbf{Known networks} --- network prefixes the agent has discovered.
    \item \textbf{Known hosts} --- host IP addresses the agent has discovered.
    \item \textbf{Controlled hosts} --- hosts the agent currently controls (always a subset of \textit{known hosts}).
    \item \textbf{Known services} --- discovered services per host (type, version, port, etc.).
    \item \textbf{Known data} --- discovered data items per host.
    \item \textbf{Known blocks} --- discovered firewall rules/blocks that prevent specific connections.
\end{enumerate}

\subsubsection{State Space}
\label{subsec:state_space}

In each NSG scenario we count only non-router hosts, non-local services, and non-log data. Under these restrictions, the scenario contains $H=11$ hosts, $N_{\mathrm{net}}=4$ networks, $N_{\mathrm{svc}}=13$ non-local service instances, $N_{\mathrm{data}}=5$ data instances, and $N_{\mathrm{block}}=20$ firewall host pairs that can be blocked. Approximating each host as being in one of three modes (unknown, known, controlled), and treating each service, data, network, and firewall-block indicator as a binary feature, the approximate environment state-space size is
\[
|S| \approx 3^{H} 2^{N_{\mathrm{svc}} + N_{\mathrm{data}} + N_{\mathrm{net}} + N_{\mathrm{block}}}
\]
\[
= 3^{11} 2^{13 + 5 + 4 + 20}
\approx 7.8 \times 10^{17}.
\]

In other environments, such as CyberBattleSim \cite{microsoft2021cyberbattlesim}, one encounters a similarly large set of underlying combinatorial configurations. However, in CyberBattleSim these environments are typically presented to learning algorithms through compact observation encodings (e.g., fixed-length feature vectors on the order of tens of dimensions). Such representations often expose global properties of the network directly, rather than requiring the agent to infer them purely through discovery, thereby simplifying the effective learning problem compared to the fully latent formulation considered here.

\subsubsection{Action Space}
The actions in NSG are parameterized by an action type and concrete arguments drawn from the agent’s current observation. In our attacker-only setting, the core action types are \texttt{ScanNetwork}, \texttt{FindServices}, \texttt{FindData}, \texttt{ExploitService}, and \texttt{ExfiltrateData}. NSG also supports defensive actions such as \texttt{BlockIP}; we do not enable defensive blocking for the attacker agents in this paper.

\begin{table*}[ht]
\centering
\footnotesize
\caption{List of NSG action types and their parameters. The attacker agents in this paper use the first five action types; \texttt{BlockIP} is a defender action.}
\begin{tabular}{l|l}
\textbf{Action type} & \textbf{Additional parameters} \\
\hline
ScanNetwork & source\_host, $\text{target\_network} \in \texttt{known\_networks}$ \\
FindServices & source\_host, $\text{target\_host} \in \texttt{known\_hosts}$ \\
FindData & source\_host, $\text{target\_host} \in \texttt{known\_hosts}$ \\
ExploitService & source\_host, $\text{target\_host} \in \texttt{known\_hosts}$, $\text{target\_service} \in \texttt{known\_services}$ \\
ExfiltrateData & source\_host, $\text{target\_host} \in \texttt{controlled\_hosts}$, $\text{data} \in \texttt{known\_data}$ \\
BlockIP (defender) & source\_host, $\text{target\_host} \in \texttt{controlled\_hosts}$, $\text{blocked\_host} \in \texttt{known\_hosts}$
\end{tabular}
\label{tab:action_space}
\end{table*}

\paragraph{Action-space catalogue.}
We can view the game’s attacker action catalogue as the set of \emph{distinct parameterized action templates} that exist in the scenario configuration, before accounting for partial observability and per-step validity constraints. Let $N$ be the number of hosts, $M$ the number of networks, $S$ the number of (non-local) service instances, and $D$ the number of data items (Section~\ref{subsec:state_space}). Counting one concrete action for each combination of an action type with its parameters gives
\begin{equation*}
|A_{\mathrm{catalog}}| = N\,M + 2N^2 + N\,S + N^2\,D.
\end{equation*}
The terms correspond to \texttt{ScanNetwork} ($N\,M$), \texttt{FindServices} ($N^2$), \texttt{FindData} ($N^2$), \texttt{ExploitService} ($N\,S$), and \texttt{ExfiltrateData} ($N^2\,D$). With $N=11$, $M=4$, $S=13$, and $D=5$, this yields $|A_{\mathrm{catalog}}| = 1034$ parameterized actions.

\paragraph{Valid actions per state and total state--action scale.}
The \emph{valid} action set presented to an agent at a given step is typically much smaller: actions are only available when their arguments are currently known (and, for some types, satisfy additional constraints such as requiring a controlled source/target), and reachability/firewall rules further restrict which host pairs can interact. If we denote by $|A(s)|$ the number of valid actions in state $s \in S$ and assume that on average there are between $10$ and $100$ valid actions per state, then
\[
7.8 \times 10^{18} \;\lesssim\; \sum_{s \in S} |A(s)| \;\lesssim\; 7.8 \times 10^{19},
\]
which is on the order of $10^{19}$ state--action pairs. As a loose worst-case upper bound, $\sum_{s\in S}|A(s)| \le |S|\,|A_{\mathrm{catalog}}| \approx (7.8\times10^{17})\times 1034 \approx 8.1\times 10^{20}$, though this bound is not tight in partially observed play.

\subsection{Firewall}
The scenario includes a router-level firewall that restricts \emph{inter-subnet} traffic between the client subnet and the server subnet. In the default configuration, only a limited set of client hosts can initiate connections to a small set of exposed servers (e.g., SMB/DB/Web), while other inter-network flows are denied. Traffic \emph{within} each /24 subnet is not filtered, so lateral movement within a subnet is unconstrained once a foothold is obtained.

For our purposes, the firewall primarily changes the \emph{reachability graph} and therefore the transition structure of the game (which actions are valid from which controlled hosts). It does not change the IP reassignment nature of our task variants; across IP assignments, the same logical constraints apply with different concrete IP values.

In this scenario (called Scenario 1 in the code), the only firewall present is on the \texttt{router}, whose global default policy is \texttt{DENY}. The INPUT chain on the FW permits hosts from the internal client subnets and server to talk directly to its own networks. No access to the router is allowed. The FORWARD chain is also \texttt{DENY} by default and contains explicit \texttt{ALLOW} rules only for flows from the five client IPs in the client network to the three main servers \verb|192.168.1.2|, \verb|192.168.1.3|, and \verb|192.168.1.4|. Any other intra-network traffic from the client network to the server network is not allowed. Connections from the server network to the client network are not allowed. Traffic from external networks into the internal subnets is not allowed. The flows explicitly allowed by this firewall are summarized in Table~\ref{tab:scenario1-fw-allow}.

\begin{table}[h]
\centering
\small
\setlength{\tabcolsep}{4pt}
\begin{tabular}{@{}ll@{}}
\hline
\textbf{Source} & \textbf{Destination} \\
\hline
Client network & Client network \\
Server network & Server network \\
Client network & \texttt{smb\_server} \\
Client network & \texttt{db\_server} \\
Client network & \texttt{web\_server} \\
Client network & Internet \\
Server network & Internet \\
\hline
\end{tabular}
\caption{Connections explicitly allowed in the Scenario of NSG.}
\label{tab:scenario1-fw-allow}
\end{table}

\subsection{Rewards}
The scenario reward function is specified in the scenario configuration. In our experiments, we use a sparse success reward with a step cost and penalties for failed actions:
\begin{verbatim}
rewards:
  success: 100
  step: -1
  fail: -10
  false_positive: -5
\end{verbatim}

\subsection{Exfiltration Scenario}
\begin{figure}[!t]
    \centering
    \includegraphics[width=0.9\linewidth]{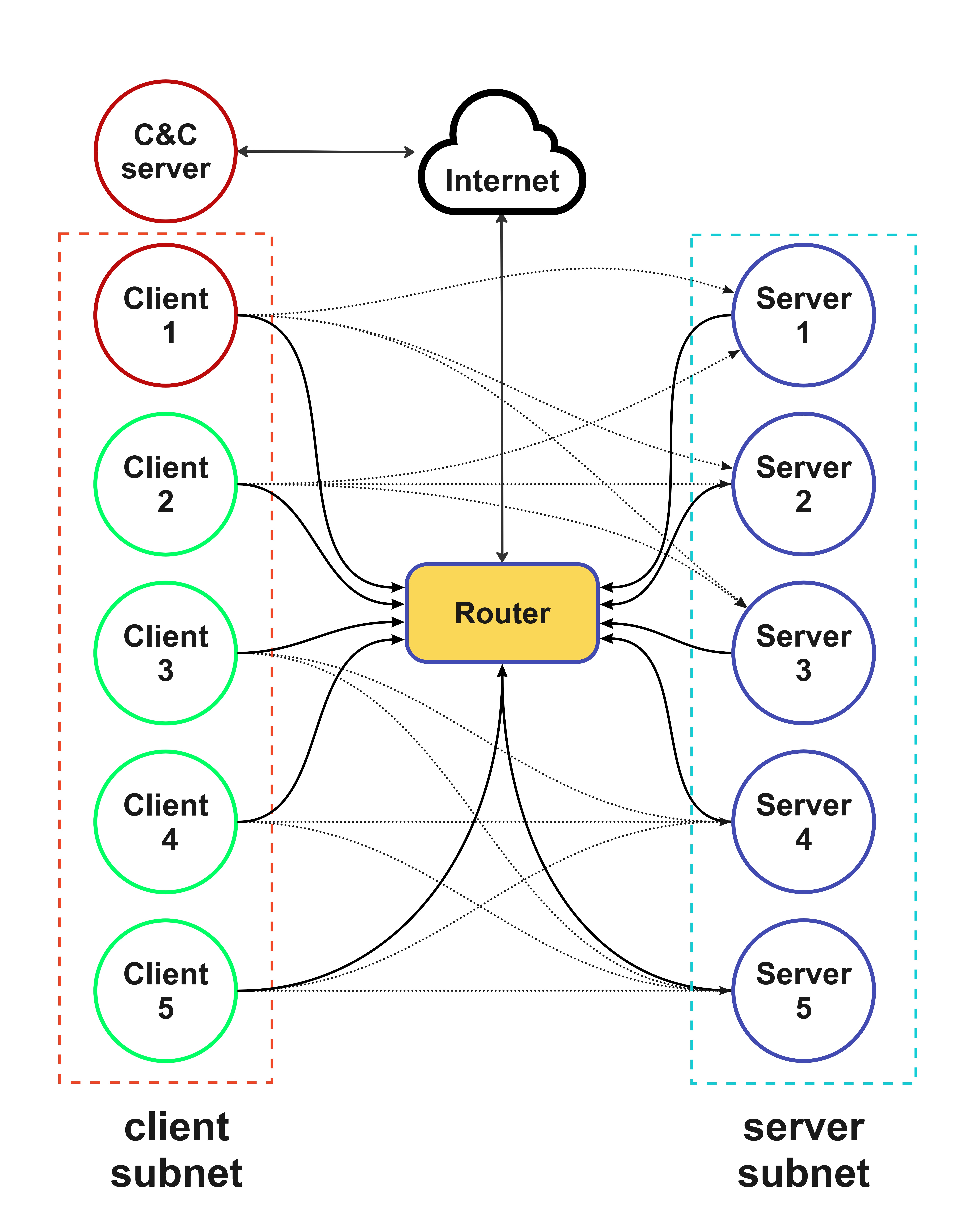}
    \caption{The network topology used in the Data exfiltration scenario. There are two local sub-networks, each consisting of 5 hosts. Hosts are reachable from each other within the same sub-networks. Access from the other sub-network is determined by firewall rules (shown as dotted lines).}
    \label{fig:task_topology}
\end{figure}

The scenario used in this work is focused on a data exfiltration attack. The exact network topology used in the scenario is shown in Figure~\ref{fig:task_topology}.

At the beginning of each episode, the agent starts with an access randomly assigned host in the client sub-network. The successful solving of the tasks consists of (i) gaining access to the server subnetwork, (ii) finding the server with the data of interest, (iii) getting access to said server, and (iv) exfiltrating data to the external C\&C server on the internet.

%%%%%%%%%%%%%%%%%%%%%%%%%%%%%%%%%%%%%%%%%%%%%
\section{Agents}

We group our agents by the main mechanism they use to handle unseen networks. This separation keeps the comparison focused on \emph{how} an agent generalizes (or fails to generalize), rather than on implementation details of a particular model.

\begin{itemize}
    \item \textbf{Traditional learning agents:} agents trained to maximize return in the training environments, typically using fixed state/action encodings (e.g., DQN/DDQN-style value learning).
    \item \textbf{LLM-based agents:} agents that use language-model reasoning and structured prompting (and in our case, an LLM+ModernBERT pipeline) to choose valid next actions from the current observation.
    \item \textbf{Generalization/adaptation agents:} agents explicitly designed to transfer across IP reassignment, either by removing address information through abstraction (conceptual Q-learning) or by learning to adapt quickly (meta-learning).
\end{itemize}

A key distinction is that \textbf{meta-learning agents perform test-time learning}: on the unseen topology they are allowed to update their parameters using a small support set of episodes before evaluation. All other agents are evaluated as fixed policies (no parameter updates) and must rely on whatever invariances, abstractions, or priors they learned during training.

\subsection{Traditional learning agents}

These agents learn a policy by optimizing return in the training environments, without any explicit mechanism for transfer to new IP assignments. A random policy is included as a no-learning reference point, and learned value-based agents (DQN and DDQN variants) are then compared under the same IP reassignment shift.

\subsubsection{Random Agent} A common baseline approach is to define an agent that selects each action independently of previous actions, which is known as random selection. In particular, for the NSG environment, the random agent selects each action independently of previous ones, following a uniform distribution over all the possible actions at a given state. 
Formally, let $S_t$ denote the environment state at time $t$, and let $\mathcal{A}(S_t)$ be the set of actions that are valid in that state, with $N_t := \lvert \mathcal{A}(S_t) \rvert$. The random policy $\pi_{\text{rand}}$ selects uniformly among valid actions:
\[
\pi_{\text{rand}}(a \mid S_t) \;=\; \begin{cases}
\frac{1}{N_t}, & a \in \mathcal{A}(S_t), \\
0, & \text{otherwise.}
\end{cases}
\]

Since it has no memory or learning mechanism, the agent may act irrationally, repeating redundant actions such as scanning networks that have already been discovered or attempting to exfiltrate data before any has been found.

\subsubsection{DQN agent}
\label{subsec:DQN_agent}
We evaluate two Deep Q-Learning (DQN) value-based attackers that learn a state--action value function over candidate actions derived from the NetSecGame environment. We consider a standard single-buffer variant and a dual-buffer variant that duplicates ``progress'' and success transitions into a compact success buffer.

\paragraph{State Representation}
The state encoding used by both DQN variants is the standard NetSecGame state without any leaked extra information. So the agent only  knows what has been discovered so far. We use a candidate-centric 12-dimensional representation, also used by the MAML agent, (Section~\ref{subsec:maml_agent}), computed for each valid action candidate from the current state:

\begin{enumerate}
  \item \textbf{Source Features (3):} number of services running on the source host, whether the source belongs to a known subnet, and source reachability ratio/centrality in the discovered graph.
  \item \textbf{Target Features (6):} normalized progress stage, whether the target is compromised, number of discovered services, whether known data exists on the host, whether the target is blocked from the source, and the target's degree in the partial topology.
  \item \textbf{Global Features (3):} owned ratio (controlled/known hosts), average degree, and graph density.
\end{enumerate}

\paragraph{First DQN Variant- \textit{Single Replay Buffer:}}
In this variant of DQN the experience tuples are stored as follows:
\begin{equation}
\label{eq:dqn_transition}
\begin{aligned}
\bigl(
&\underbrace{\mathbf{s}\in\mathbb{R}^{12}}_{\text{candidate vector}},
\underbrace{a\in\{0,\ldots,4\}}_{\text{action head}},
 r,\\
&\underbrace{\mathbf{S}'\in\mathbb{R}^{N'\times 12}}_{\text{next-step candidate matrix}},
 d\bigr),
\end{aligned}
\end{equation}

where $r$ is the reward, $d$ the terminal flag, and $\mathbf{S}'$ stacks all valid next-step candidate embeddings (empty if terminal).

\textbf{Second DQN Variant- \textit{Dual Replay Buffer:}} In this variant, besides a uniform buffer $\mathcal{D}_{\text{all}}$, we maintain a compact ``success'' buffer $\mathcal{D}_{\text{succ}}$ that duplicates transitions when progress or terminal success occurs.

\paragraph{Q-Network and Action Selection}
% The DQN agent's policy network consists of three fully connected layers, where the input layer receives the 12-dimensional state vector. Given a batch of candidate vectors, the network outputs five Q-values per $N$ candidates, each corresponding to one of the five action types in NetSecGame: \textbf{ScanNetwork, FindServices, FindData, ExploitService, ExfiltrateData}. We apply $\epsilon$-greedy selection over the flattened $N\times 5$ table: with probability $\epsilon$ we sample a random (candidate, action) pair; otherwise we take the global argmax.

The DQN agent employs a three-layer fully connected neural network that takes as input a 12-dimensional candidate feature vector. At each timestep, the agent first enumerates the set of valid parameterized actions provided by the environment (e.g., scanning a specific subnet or exploiting a specific service). Redundant actions, such as previously executed scans, are filtered prior to evaluation.

Let $\mathcal{A}(s)$ denote the set of valid parameterized actions in state $s$, and let $N = |\mathcal{A}(s)|$. For each $a_i \in \mathcal{A}(s)$, we construct a 12-dimensional feature vector $x_i$ that encodes both the current state and the contextual information associated with that specific action. These candidate vectors are stacked into a batch
\[
X = \{x_1, x_2, \dots, x_N\}.
\]

The policy network $Q_\theta(\cdot)$ maps each candidate vector $x_i$ to five Q-values:
\[
Q_\theta(x_i) \in \mathbb{R}^5,
\]
corresponding to the five high-level action refinements in NetSecGame: 
\texttt{ScanNetwork}, \texttt{FindServices}, \texttt{FindData}, \texttt{ExploitService}, and \texttt{ExfiltrateData}.

This produces an $N \times 5$ Q-value table, where each row corresponds to a single valid parameterized action context. Importantly, action selection is restricted to this dynamically constructed valid action set, ensuring that only feasible parameterizations are considered.

We apply $\epsilon$-greedy selection over the flattened $N \times 5$ Q-value table. With probability $\epsilon$, the agent samples a random valid $(i, j)$ pair, where $i$ indexes the candidate action and $j \in \{1,\dots,5\}$ indexes the action refinement head. Otherwise, the agent selects the global maximizer:
\[
(i^*, j^*) = \arg\max_{i,j} Q_\theta(x_i)_j.
\]

The executed action inherits its parameters directly from the selected candidate $a_{i^*}$, ensuring that all actions taken by the agent are valid within the environment.

\paragraph{Training and Testing Procedure}
Two training/testing configurations are used to distinguish \emph{within-topology learning} from \emph{transfer under IP reassignment}. Configuration~\#1 measures generalization by training across five IP assignments and evaluating on a sixth unseen IP assignment. Configuration~\#2 is a sanity check that measures whether the agent can learn effectively when training and evaluation use the same topology.

\textbf{Configuration \#1 (train on five IP assignments, test on unseen IP assignment):}
For each of the five training topologies, 1,000 training episodes are run with $\epsilon$-greedy exploration. At each step the agent:
\begin{enumerate}
  \item Enumerates valid actions and builds the $N\!\times\!12$ candidate matrix;
  \item Selects a (candidate, action) pair via $\epsilon$-greedy over the $N\!\times\!5$ Q-table;
  \item Steps the environment and pushes the transition with the next step’s candidate matrix into replay;

  \textbf{\textit{In the Dual-buffer:}} transitions are mirrored to $\mathcal{D}_{\text{succ}}$ whenever a progress or terminal success occurs;
  \item Samples a minibatch once the buffer exceeds a minimum size and performs a temporal difference (TD) update using a one-step bootstrapped target;

  \textbf{\textit{In the Dual-buffer:}} minibatches are formed with a mix of success and all samples for the TD update.
\end{enumerate}

For final evaluation, training is frozen and the greedy policy (no exploration or updates) is deployed on a previously unseen topology for 500 episodes.

\textbf{Configuration \#2 (train and test on the same topology):}
The agent is trained on \textbf{one} topology using the same steps as Configuration~\#1. After training for 1,000 episodes, training is frozen, the environment is reset on the same topology, and the agent is evaluated with 500 episodes.

\paragraph{Results}
DQN results are reported in Section~\ref{sec:results_traditional}.

\subsubsection{DDQN agent}
The Double Deep Q-Network (DDQN) agent extends the Deep Q-Network (DQN) by maintaining a separate target network, which reduces Q-value overestimation and stabilizes learning. In NetSecGame, the observation is provided as a structured JSON object (serialized as a string) and is first converted into a dense embedding. The Q-network then predicts Q-values for the full discrete action set $\mathcal{A}$. At the start of each episode, the environment provides $|\mathcal{A}|$ as well as a per-step valid-action mask, which constrains action selection; therefore, the agent operates with environment-provided action-space structure (i.e., a ``white-box'' action interface).

\paragraph{State Representation}
The DDQN agent encodes the serialized JSON state using the Qwen3-Embedding-0.6B~\cite{qwen3embedding} encoder through the Sentence Transformers library~\cite{reimers-2019-sentence-bert}, producing a fixed-dimensional vector $\mathbf{s}\in\mathbb{R}^d$.

\paragraph{Q-network and Action Selection}
Given $\mathbf{s}$, the Q-network outputs a vector $Q(s,\cdot)\in\mathbb{R}^{|\mathcal{A}|}$. We use an MLP with two hidden layers (1024 and 512 units) followed by a linear output layer of size $|\mathcal{A}|$. During inference, invalid actions are masked out and the agent selects the greedy valid action.

\begin{figure*}[!t]
  \centering
  \resizebox{\textwidth}{!}{%
  \begin{tikzpicture}[
    node distance=0.25cm and 0.7cm,
    every node/.style={draw, rectangle, align=center, font=\scriptsize, minimum width=2.2cm, rounded corners=2pt},
    >=Stealth
  ]
    \node (state) {Serialized JSON \\ observation};
    \node (embed) [right=of state, fill=green!10] {Qwen3-Embedding-0.6B \\ state encoder \\ $\mathbf{s}\in\mathbb{R}^d$};

    \node (mlp) [right=of embed, fill=yellow!15] {MLP Q-network \\ ($d\rightarrow1024\rightarrow512\rightarrow|\mathcal{A}|$)};
    \node (qvec) [right=of mlp, fill=purple!10] {$Q(s,\cdot)$ vector \\ over full action set};

    \node (valid) [below=0.8cm of qvec, fill=blue!8] {Valid-action mask \\ from environment};
    \node (select) [right=of qvec, fill=orange!10] {Greedy selection \\ $\arg\max$ over valid actions};

    \node (target) [below=0.8cm of mlp, fill=purple!5] {Target network \\ (DDQN stability)};

    \draw[-{Stealth}] (state) -- (embed);
    \draw[-{Stealth}] (embed) -- (mlp);
    \draw[-{Stealth}] (mlp) -- (qvec);
    \draw[-{Stealth}] (qvec) -- (select);
    \draw[-{Stealth}] (valid) -- (select);
    \draw[-{Stealth}] (mlp) -- (target);
  \end{tikzpicture}%
  }
  \caption{DDQN agent architecture: JSON observations are embedded, mapped to a Q-value vector over the full action set, and the greedy action is selected after masking to currently valid actions (with a delayed target network for stability).}
  \label{fig:ddqn_arch}
\end{figure*}
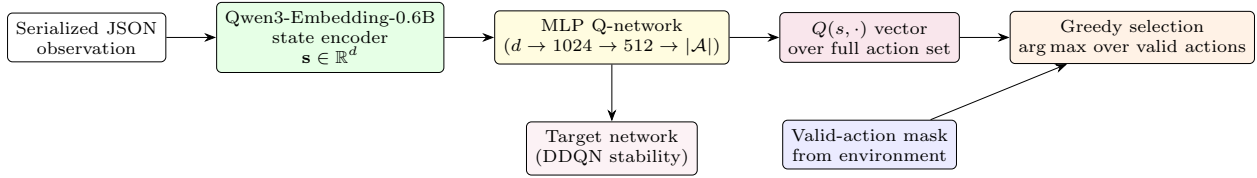

Figure~\ref{fig:ddqn_arch} summarizes the embedding and decision pipeline used by the DDQN agent.

\paragraph{Training and Testing Procedure}
The agent is trained sequentially on five network topologies until convergence and evaluated on a sixth, unseen topology. As in our DQN baseline (Section~\ref{subsec:DQN_agent}), we use a dual replay buffer for experience storage and an $\epsilon$-greedy exploration policy.

Because NetSecGame rewards are sparse, we add a simple intrinsic reward term that encourages state changes and discourages redundant actions. The reward at time step $t$ is

\begin{equation}
r_t = \frac{1}{10} \left( R_{env} +
\begin{cases}
50, & \text{if } s_t \neq s_{t+1} \\
-50, & \text{if } s_t = s_{t+1}
\end{cases} \right),
\end{equation}

where $R_{env}$ is the environment reward and $s_t$ is the state at time $t$.

\paragraph{Results}
DDQN results are reported in Section~\ref{sec:results_traditional}.

\subsection{LLM-based agents}

\subsubsection{ReAct agent} 
\label{sec:react_agent}

Recent studies have shown that LLM-based agents have outstanding capabilities in penetration testing scenarios, largely thanks to the models’ ability to synthesize reconnaissance results, propose exploit sequences, and reason over conditional action spaces in natural language~\cite{happe2023getting,moskal2023llms}. These works demonstrate that even few-shot prompting with ReAct-style chains can recover meaningful action plans, while fine-tuning enables longer-horizon tactical reasoning when paired with structured thinking (e.g., generating JSON-formatted commands). The flexibility of LLMs also allows them to ingest textual rules and respond in domain-specific formats, which is why we explored the ReAct architecture alongside encoder-based hybrids in this study.

\paragraph{Architecture}
The ReAct agent~\cite{rigaki_out_2024,RIGAKI2026129987} combines reasoning and action selection within a two-stage prompting framework inspired by the ReAct~\cite{yao_react_2022} architecture. 

In the first stage, the language model analyzes the current network state provided by the environment, identifies relevant objects such as hosts, networks, and services, and reasons about the possible actions that can be taken (see appendix A.1 of~\cite{rigaki_out_2024} for the complete prompt).

In the second stage, it selects the most appropriate next action and its parameters, organizing them in a well-defined JSON structure (see appendix A.2 of~\cite{rigaki_out_2024} for the complete prompt).

Each prompt includes the environment rules, the textual representation of the current state, examples of valid actions, and a \emph{prompt window} of recent actions. This prompt window contains the last $k$ actions taken by the agent along with their outcomes, allowing the model to avoid repeating ineffective or redundant actions and to maintain local context across decision steps.
While $k$ is a tunable parameter representing the length of previous actions, we set $k=10$ for all experiments in this work.
Importantly, the agent uses \emph{no persistent memory between episodes} (i.e., no cross-episode state is carried over). The underlying LLM is also \emph{not trained online}: its parameters remain fixed during evaluation, with no gradient updates or learning from past episodes.

We employ gpt-oss-120b~\cite{openai2025gptoss120bgptoss20bmodel}, a state of the art open weights model, as the underlying LLM. The model supports different levels of reasoning effort, agentic capabilities, and structured outputs.

\paragraph{Results}
ReAct results are reported in Section~\ref{sec:results_llm}.

%\begin{table}[t]
%\centering
%\caption{Episode Summary Statistics for GPT-OSS-120B on Topology 6}
%\label{tab:summary}
%\footnotesize
%\setlength{\tabcolsep}{3pt}
%\resizebox{\columnwidth}{!}{%
%\begin{tabular}{@{}lrrr@{}}
%\toprule
%\textbf{Metric} & \textbf{All Episodes} & \textbf{Wins} & \textbf{Losses} \\
%\midrule
%Count & 263 & 250 & 13 \\
%Win Rate (\%) & 95.1 & --- & --- \\
%Avg Steps & 31.2 & 27.8 & 96.7 \\
%Avg Total Reward & 63.9 & +72.2 & -97.5 \\
%Min Steps & 8 & 8 & 91 \\
%Max Steps & 100 & 91 & 100 \\
%\bottomrule
%\end{tabular}%
%}
%\end{table}

\paragraph{Failure analysis}
ReAct failure-mode and robustness analyses are reported in Section~\ref{sec:results_llm}.

\subsubsection{LLM-BERT agent}

The LLM-BERT agent combines large language model reasoning with efficient BERT-based action generation. This hybrid architecture addresses the computational cost of pure LLM agents while maintaining reasoning capabilities for complex decision making in cybersecurity scenarios.

\paragraph{Agent Architecture}

Building upon the ReAct framework described in Section~\ref{sec:react_agent}, this agent operates in a modified pipeline that retains the initial reasoning phase but replaces the generative action selection with specialized discriminative models. The process consists of three stages:

\textbf{Stage 1 (Reasoning):} This stage mirrors the first phase of the ReAct agent. An LLM (gpt-oss-120b) processes the current environment state—including controlled hosts, known networks, and services—to produce a comprehensive structured situational analysis. However, rather than directly outputting an action, this analysis serves as the contextual input for the subsequent BERT modules.

\textbf{Stage 2 (Action Classification):} A fine-tuned ModernBERT-large model (parameter size: 395M) configured as \texttt{AutoModelForSequenceClassification} processes the complete prompt from Stage 1 to predict the action type. The classifier selects from five possible actions: \texttt{ScanNetwork}, \texttt{ScanServices}, \texttt{ExploitService}, \texttt{FindData}, or \texttt{ExfiltrateData}.

\textbf{Stage 3 (Parameter Generation):} Instead of a single general-purpose MLM, this stage employs five specialized ModernBERT-base models (149M parameters each), with one model dedicated to each action type. Here, \emph{MLM} denotes a \emph{masked language model}: a BERT-style model trained to predict intentionally hidden tokens (\texttt{[MASK]}) from surrounding context. Each specialized model is configured as \texttt{AutoModelForMaskedLM} and trained exclusively on examples of its corresponding action. A router with LRU caching manages these specialized models, loading them on-demand and maintaining at most three models in GPU memory to optimize resource utilization. The selected specialized model fills a JSON action template with appropriate parameters. The template contains \texttt{[MASK]} tokens at parameter positions, which the model replaces. The number of \texttt{[MASK]} tokens for each parameter was inferred from the dataset to match the maximum observed token length for that parameter type. The action templates are structured as follows:

\begin{lstlisting}[
  basicstyle=\small\ttfamily,
  breaklines=true,
  breakatwhitespace=true,
  xleftmargin=0pt,
  columns=flexible
]
{
  "ExfiltrateData": {
    "action": "ExfiltrateData",
    "parameters": {
      "target_host": "[MASK].[MASK].[MASK].[MASK]",
      "data": {
        "owner": "[MASK][MASK][MASK]",
        "id": "[MASK][MASK][MASK][MASK][MASK][MASK]"
      },
      "source_host": "[MASK].[MASK].[MASK].[MASK]"
    }
  }
}
\end{lstlisting}

When an action requires fewer tokens than the maximum allocated, the model generates \texttt{[PAD]} tokens to maintain a consistent output structure. Special token normalization ensures consistency with the training data, and post-processing removes \texttt{[PAD]} and \texttt{[UNK]} tokens from the final output.

\paragraph{Fine-tuning Process}

% \paragraph{Dataset Generation}

\begin{figure}[!t]
    \centering
    \includegraphics[width=1\linewidth]{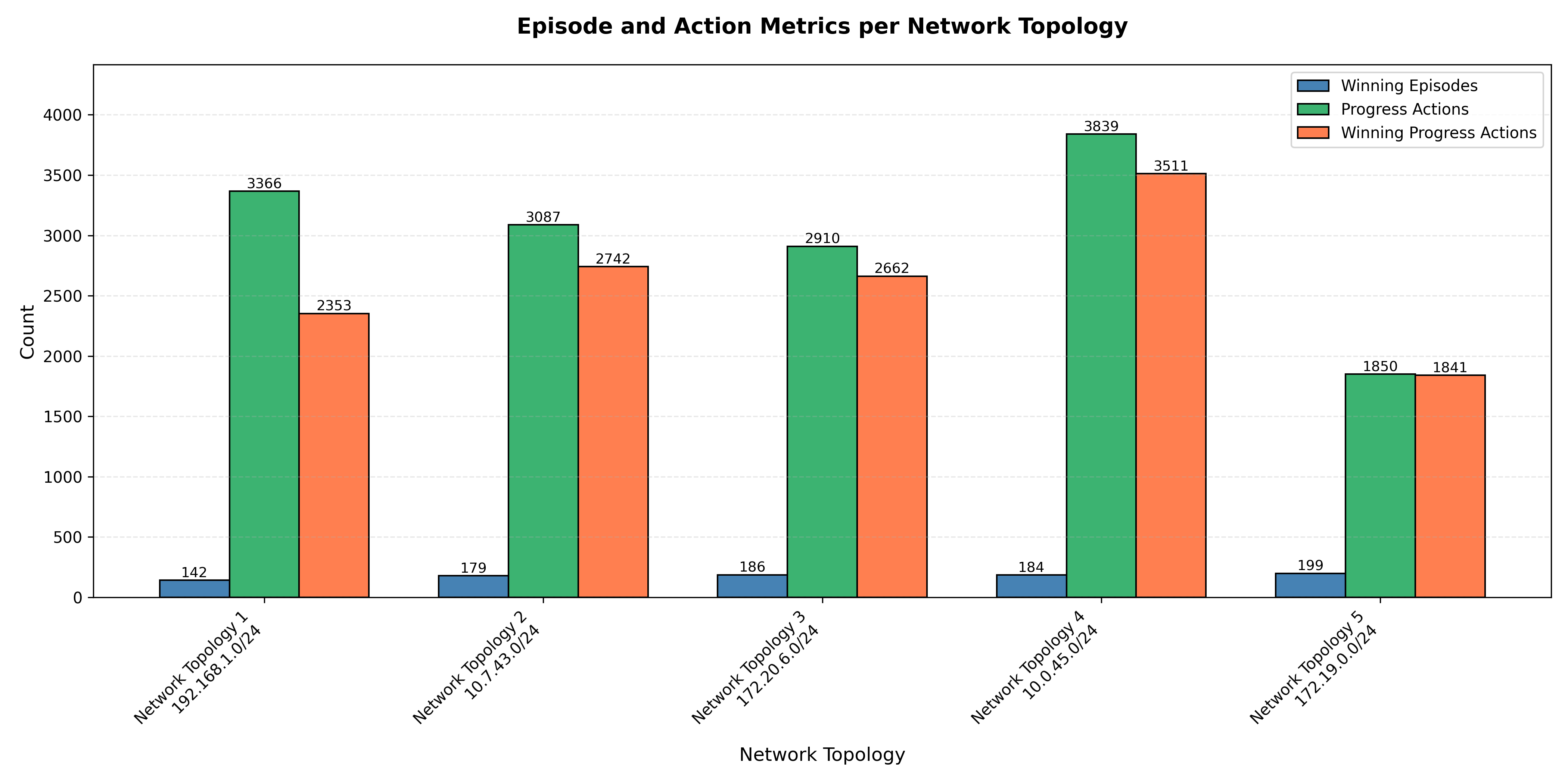}
    \caption{Dataset distribution across topologies. Win values represent successful episodes, Progress actions are state changing valid actions, and Winning Progress actions are progress actions from successful episodes.}
    \label{fig:dataset_distribution}
\end{figure}

To train the \emph{LLM-BERT} pipeline (Stage~2 classifier and Stage~3 MLM parameter-fillers), we first generated a supervised dataset by running the ReAct agent (GPT-OSS-120b) across the five training topologies and logging its (prompt, action) traces. Figure~\ref{fig:dataset_distribution} reports the composition of this \textbf{LLM-BERT fine-tuning dataset} across topologies, including counts of successful episodes (wins), all progress actions, and progress actions restricted to successful episodes.

Progress actions are defined as valid actions that successfully change the environment state, distinguishing them from invalid or redundant actions that have no effect.

The data collection yielded varying amounts of useful actions across topologies: Network Topology 1 (192.168.1.0/24) provided 142 winning episodes with 3,366 progress actions (2,353 from winning episodes), Network Topology 2 (10.7.43.0/24) contributed 179 wins with 3,087 progress actions (2,742 winning), Network Topology 3 (172.20.6.0/24) had 186 wins with 2,910 progress actions (2,662 winning), Network Topology 4 (10.0.45.0/24) achieved 184 wins with 3,839 progress actions (3,511 winning), and Network Topology 5 (172.19.0.0/24) performed best with 199 wins and 1,850 progress actions (1,841 winning).

\paragraph{Model Training}

\textbf{Stage 2 Classifier Training:} The ModernBERT-large model (395M parameters) was fine-tuned for action classification using a balanced dataset of 7,880 examples, with 1,576 examples per action type (20\% each). This balanced subset ensures the classifier learns to recognize all action types equally, preventing bias toward more frequent actions. Training configuration: 7 epochs, batch size of 1 with gradient accumulation over 4 steps (effective batch size of 4), learning rate of $5 \times 10^{-5}$, and a 90/10 train-validation split. The model was trained on prompts labeled with their corresponding action types executed by the LLM ReAct agent.

\textbf{Stage 3 MLM Training:} Five specialized masked language models (ModernBERT-large, 395M parameters each) were trained, one for each action type. The MLM models were trained on the full dataset of 13,109 examples, which maintains the natural distribution of actions in gameplay: ScanServices (23.70\%), FindData (25.62\%), ExploitService (22.47\%), ExfiltrateData (16.18\%), and ScanNetwork (12.02\%). Each model was trained exclusively on examples of its corresponding action. For example, the ExploitService MLM was trained only on ExploitService examples (2,946 examples), the ScanNetwork MLM only on ScanNetwork examples (1,576 examples), and so forth. Action parameters were replaced with \texttt{[MASK]} tokens in the input, while the complete JSON actions served as targets. Training used variable epochs optimized per action type: FindData (10 epochs), ScanServices (10 epochs), ExploitService (6 epochs), ExfiltrateData (13 epochs), and ScanNetwork (6 epochs). Other training parameters: batch size of 1 with 16-step gradient accumulation, learning rate of $2 \times 10^{-5}$, and a 90/10 train-validation split. Special token handling included vocabulary resizing and space normalization to ensure consistency between training and inference.

The specialized MLM approach enables efficient inference by allowing each model to learn action-specific parameter patterns more deeply than a single general-purpose model. 
ModernBERT's smaller size (compared to gpt-oss-120b) allows rapid action generation while maintaining the reasoning capabilities provided by the initial LLM analysis. This architecture balances computational efficiency with decision making sophistication, crucial for real-time cybersecurity applications.

\paragraph{Results}
LLM-BERT results and failure analyses are reported in Section~\ref{sec:results_llm}.

\subsection{Generalization/Adaptation Agents}

These agents include an explicit mechanism intended to work across unseen IP reassignment. Two approaches are studied: (i) \emph{abstraction}, which removes concrete address information from the agent’s internal representation, and (ii) \emph{adaptation}, where the agent is trained to update quickly with a small amount of experience in the unseen network.

\subsubsection{Conceptual Agent}
% What is the conceptual agent? Motivation
The conceptual agent is a modified Q-learning agent designed to operate with \textit{concepts} rather than static values from the environment, such as IP addresses, network addresses, or port numbers. The motivation behind the conceptual agent was to provide an abstraction mechanism that enables the agent to operate in new network topologies by transferring the actions learned on concepts from previous topologies. The \textit{concepts} used by the agent aims to translate values, such as an IP address, into how a human would perceive them, for example, a \textit{host with a port 22/TCP open}.

% How does it work
The essence of the conceptual agent is quite simple: it is a set of fixed heuristics that map the different parts of the state from the environment into how a human would remember them. If a host is found with port 22/TCP open, then the agent remembers it by assigning the string name \textit{host0\_port22/TCP} instead of its IP address. The entire agent then deals only with these \textit{concepts} for learning. Since the entire state is transformed into a concept-state, the Q-learning agents continue to create the Q-table, choose an action, and so on, using the concept-state.

Since the conceptual agents operate internally with concept-states, the actions they create use concepts instead of real values, such as IP addresses. After a concept-action is chosen, it undergoes another translation into a real action in this environment, which is ultimately used to play.

\begin{figure}
    \centering
    \includegraphics[width=1\linewidth]{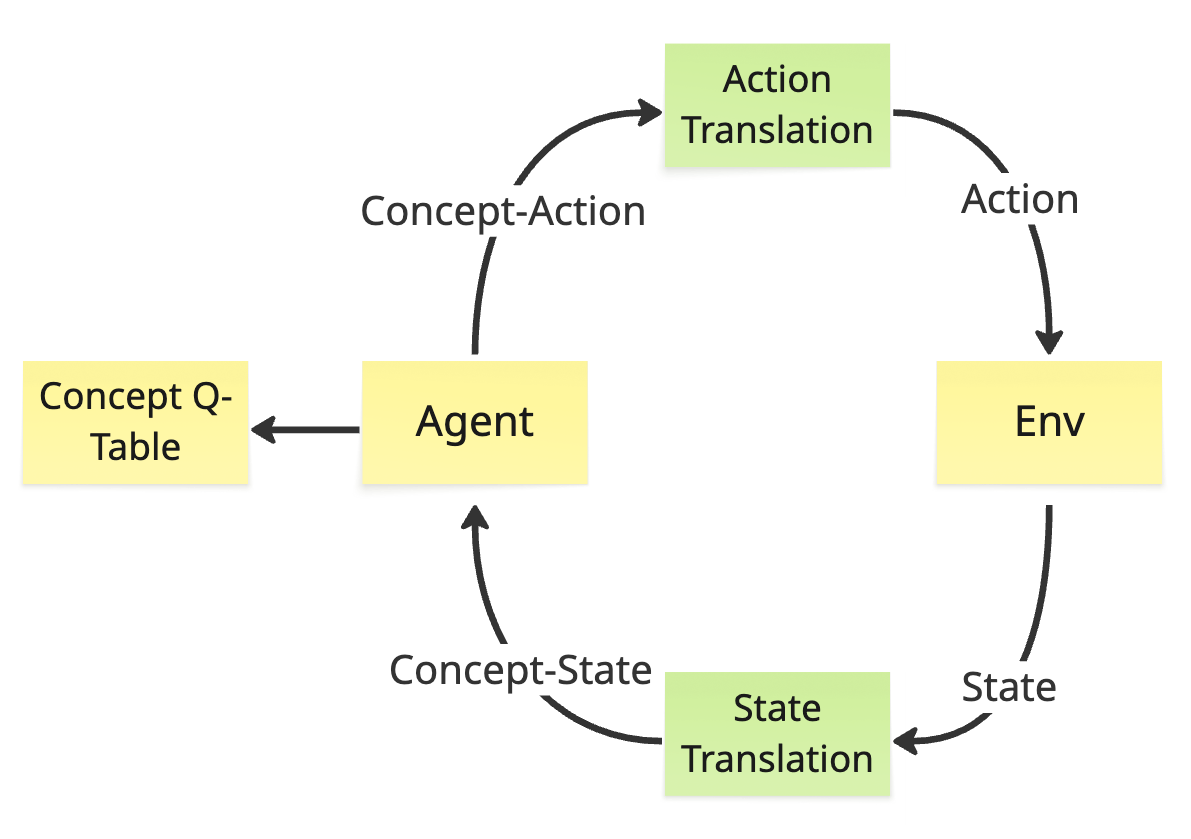}
    \caption{Diagram of Operation of the Conceptual Agent. The translations from state to concept-state and from concept-action to action are done as a wrapper. The agent operates on the new state without information about the original state.}
    \label{fig:conceptual_agent_diagram}
\end{figure}

% Architecture
The architecture diagram of the conceptual agent is shown in Figure~\ref{fig:conceptual_agent_diagram}. The translations act as wrappers, allowing the agent (actually \textit{any} agent) to receive and work on a modified state that now has concepts instead of IPs, networks, and services. The core of the conceptual agent is more about the translation mechanisms than the agent's algorithm.

% Which are the concepts translations and how they are done.
The translation of values from the environment to concepts is achieved through a general heuristic that attempts to mimic human knowledge; thus, in a sense, it can be considered an expert system.

\paragraph{State Abstraction}

A concrete network state is denoted
\[
  s = (H_c, H_k, S, D, N, B),
\]
where $H_c \subseteq H_k$ are controlled and known hosts, $S : H_k \to \mathcal{P}(\text{Service})$ gives known services per host, $D : H_k \to \mathcal{P}(\text{Data})$ gives known data per host, $N$ is the set of known networks, and $B \subseteq H_k \times H_k$ encodes known blocks. This description depends on concrete addresses (e.g. IPs).

\paragraph{Host and service concepts.}
The conceptual agent defines an abstraction
\[
  \alpha_H : H_k \to C_H
\]
from concrete hosts to host–concepts $ C_H$. Each host is first classified by coarse properties such as internal vs.\ external and controlled vs.\ uncontrolled. Then this is refined using its observable service set $S(h)$ (e.g.\ which ports or service types it offers). Two hosts that differ only by address but have the same role (e.g.\ internal controlled web server with SSH) are mapped to concepts of the same structural form. Thus, $\alpha_H$ removes address information while preserving behaviourally relevant structure.

\paragraph{Data, networks, and conceptual state.}
Data and networks are lifted through $\alpha_H$ as well. For data, the conceptual mapping is
\[
  \widetilde{D}(\alpha_H(h)) = D(h),
\]
so the data is indexed by host–concepts instead of concrete hosts. For networks, an abstraction
\[
  \alpha_N : N \to C_N
\]
maps each internal network $N$ to a network–concept $C_N$ that captures structural features (e.g.\ how many known hosts it contains), again ignoring its concrete network prefix. The resulting concept–state is
\[
  \tilde{s} = (\tilde{H}_c, \tilde{H}_k, \tilde{S}, \tilde{D}, \tilde{N}, \tilde{B}),
\]
where $\tilde{H}_k = \alpha_H(H_k)$, $\tilde{H}_c = \alpha_H(H_c)$, $\tilde{S}$ and $\tilde{D}$ are defined via $\alpha_H$, $\tilde{N} = \alpha_N(N)$, and $\tilde{B}$ is obtained by applying $\alpha_H$ to both endpoints of each pair in $B$. Q--learning is then performed on $\tilde{s}$, yielding policies that are invariant to concrete addressing.

The heuristics used to translate concrete network states into concept–states are summarized in Table~\ref{tab:concept-heuristics}. Each row specifies how a particular type of element in the raw state (hosts, services, data, networks, and block relations) is mapped to an abstract concept symbol based only on structural properties (e.g., internal/external location, control status, service surface, or host counts). This table thus makes explicit which concrete details are retained in the conceptual representation and which are discarded as irrelevant for learning.

\begin{table*}[!t]
\centering
\scriptsize
\caption{Conceptual Agent: heuristic mapping from concrete state elements to address-independent concepts. Abbreviations: ext.=external, int.=internal, ctrl.=controlled, svc.=non-local service, $H(n)$=\#known hosts in network $n$.}
\renewcommand{\arraystretch}{1.05}
\setlength{\tabcolsep}{4pt}
\begin{tabular}{p{3.2cm}p{3.6cm}p{5.6cm}}
\toprule
\textbf{Concrete element} & \textbf{Condition} & \textbf{Concept representation} \\
\midrule
Host $h$ & ext. & \texttt{external\_k} \\
Host $h$ & int. \& ctrl. & \texttt{host\_k} \\
Host $h$ & int. \& $\neg$ctrl. & \texttt{unknown\_k} \\
Host $h$ + $S(h)$ & svc. discovered & \texttt{host\_k\_p\_1\_p\_2\ldots} (service-derived tags $p_i$) \\
Data $d$ on host $h$ & $d$ discovered & attach $d$ to host concept $c_h$ (i.e., $\tilde{D}(c_h)\ni d$) \\
Network $n$ & int.; count $H(n)$ & \texttt{net\_k\_}\,$H(n)$\texttt{hosts} \\
Block $(h_1\!	o\!h_2)$ & block discovered & $(c_{h_1},c_{h_2})$ \\
\bottomrule
\end{tabular}
\label{tab:concept-heuristics}
\end{table*}

\paragraph{Action Translation}
When the conceptual agent selects an action, it does so entirely in the abstract space: sources, targets, and networks are represented by host–concepts and network–concepts rather than concrete addresses. To actually interact with the environment, each conceptual action is instantiated into a concrete one by applying inverse mappings (denoted $\psi_H$ for hosts and $\psi_N$ for networks) that map a concept back to an appropriate concrete host or network, while leaving non-structural payloads (such as service identifiers or data items) unchanged. In this way, learning and planning happen at the conceptual level, but execution always uses valid, concrete parameters.

The concrete instantiation rules for each action type are summarized in Table~\ref{tab:action-heuristics}. Each row specifies which conceptual parameters are used (host concepts, network concepts, data, services) and how $\psi_H$ and $\psi_N$ are applied to obtain a concrete source host, target host, or target network, optionally with additional constraints (e.g.\ source must be controlled). This makes explicit how every conceptual decision is converted into a valid low-level action in the underlying network environment.

\begin{table*}[ht]
\centering
\caption{Conceptual Agent: Heuristic translation of conceptual actions into concrete environment actions via the inverse abstractions $\psi_H$ (hosts) and $\psi_N$ (networks).}
\renewcommand{\arraystretch}{1.2}
\begin{tabular}{p{3cm}p{5cm}p{7cm}}
\hline
\textbf{Conceptual action} & \textbf{Concept-level parameters} & \textbf{Translation to concrete action} \\
\hline
Exploit service &
Source host concept $\tilde{h}_{\mathrm{src}}$, target host concept $\tilde{h}_{\mathrm{tgt}}$, service descriptor $\tilde{s}$ &
$h_{\mathrm{src}} = \psi_H(\tilde{h}_{\mathrm{src}})$ chosen among controlled hosts, $h_{\mathrm{tgt}} = \psi_H(\tilde{h}_{\mathrm{tgt}})$ among known hosts; $\tilde{s}$ is kept unchanged, yielding a concrete service to exploit $(h_{\mathrm{src}}, h_{\mathrm{tgt}}, \tilde{s})$.\\

Find services &
Source concept $\tilde{h}_{\mathrm{src}}$, target concept $\tilde{h}_{\mathrm{tgt}}$ &
$\psi_H$ is applied to obtain a concrete controlled source $h_{\mathrm{src}}$ and a concrete known target $h_{\mathrm{tgt}}$, instantiating a service-discovery scan from $h_{\mathrm{src}}$ to $h_{\mathrm{tgt}}$. \\

Find data &
Source concept $\tilde{h}_{\mathrm{src}}$, target concept $\tilde{h}_{\mathrm{tgt}}$ (both representing controlled hosts) &
Both concepts are mapped via $\psi_H$ restricted to controlled hosts, producing concrete controlled machines $(h_{\mathrm{src}}, h_{\mathrm{tgt}})$ and a data-search action $h_{\mathrm{src}} \to h_{\mathrm{tgt}}$. \\

Exfiltrate data &
Source concept $\tilde{h}_{\mathrm{src}}$, target concept $\tilde{h}_{\mathrm{tgt}}$, data item $\tilde{d}$ (all on controlled hosts) &
$\psi_H$ yields concrete controlled hosts $h_{\mathrm{src}}$ and $h_{\mathrm{tgt}}$; the data identifier $\tilde{d}$ is preserved, resulting in a concrete exfiltration $(h_{\mathrm{src}}, h_{\mathrm{tgt}}, \tilde{d})$. \\

Scan network &
Source host concept $\tilde{h}_{\mathrm{src}}$, network concept $\tilde{n}$ &
$h_{\mathrm{src}} = \psi_H(\tilde{h}_{\mathrm{src}})$ is a concrete controlled host; the target network is instantiated as $n = \psi_N(\tilde{n})$, choosing one concrete network consistent with the concept, and a network scan $h_{\mathrm{src}} \to n$ is issued. \\
\hline
\end{tabular}

\label{tab:action-heuristics}
\end{table*}

\paragraph{Heuristic Selection of Actions to Improve Performance}
To keep training time manageable in large, partially known networks, the conceptual agent prunes the action space by applying a set of domain-specific filters when generating candidate actions. These filters remove actions that are provably impossible (e.g., blocked by known firewalls), structurally redundant (e.g., re-scanning or re-exfiltrating the same target). The main heuristics are listed in Table~\ref{tab:action-filters-short}, which shows how each filter is defined and why it is beneficial for both realism and sample efficiency. In particular, NetSecGame generates on each host a file with the list of all the actions taken on that host by all agents, which is ignored for the purposes of exfiltration.

\begin{table}[t]
\centering
\footnotesize
\caption{Conceptual Agent: main filters for reducing the candidate action set while preserving informative options for learning.}
\setlength{\tabcolsep}{3.5pt}
\renewcommand{\arraystretch}{1.08}
\begin{tabularx}{\columnwidth}{p{0.42\columnwidth}X}
\toprule
\textbf{Filter} & \textbf{Description} \\
\midrule
Internal-only + firewall-aware & Keep only internal actions; drop actions that traverse a known blocked link. \\
No repeats (same concept-action) & Do not propose an action if the same (source, target, type) concept-action was already attempted. \\
Scan only when svc. unknown & Propose service-discovery only for hosts with unknown services. \\
No exploit on ctrl./self/local & Exploit only uncontrolled targets; disallow self-target; ignore local-only services. \\
FindData only if needed & Search for data only on controlled internal hosts without known data. \\
Exfiltrate only new relevant data & Exfiltrate only non-duplicate non-log data to another controlled host (no self-exfiltration). \\
\bottomrule
\end{tabularx}
\label{tab:action-filters-short}
\end{table}

\paragraph{Reward Engineering}
The conceptual Q-learning attacker performs explicit reward engineering to turn the sparse environment feedback into a dense, shaped learning signal. After every interaction, the raw environment reward and termination reason are passed through the method \texttt{recompute\_reward}, which overwrites the reward according to high-level security outcomes: if the episode ends in detection or failure (\texttt{AgentStatus.Fail}), the agent receives a strong negative reward of $-1000$; if it successfully compromises the target (\texttt{AgentStatus.Success}), it receives a strong positive reward of $+1000$; if the episode terminates by timeout, it receives a penalty of $-100$. To discourage unproductive probing and loops, the agent tracks the previous state and assigns an internal reward of $-100$ whenever the current state is identical to the previous one, and a small step cost of $-1$ otherwise.

\paragraph{Limitations and Problems}
While the conceptual abstraction significantly reduces the size of the state and action spaces, it also introduces several limitations. The key issue is that the mapping from concrete states/actions to concepts is many-to-one, whereas the grounding of a chosen concept-action back into the concrete environment is one-to-many. When the agent selects an abstract action, this must be instantiated as a concrete action on a specific IP address. If multiple concrete hosts are compatible with the same concept, the grounding mechanism chooses one of them, typically at random.

\paragraph{Training and testing procedure}
The conceptual agent is trained sequentially on five IP-reassignment variants of the same scenario using the engineered reward described above and a 100-step horizon. Evaluation is performed on a sixth, unseen IP assignment. To characterize variability due to stochastic grounding of concepts to concrete hosts, we repeat testing across multiple random seeds.

Conceptual agent results are reported in Section~\ref{sec:results_adapt}.

\subsubsection{MAML agent}
\label{subsec:maml_agent}

We implement an MAML-based attacker agent for NetSecGame. The agent learns a policy initialization that can adapt to a new network layout in a few gradient steps using task-specific (inner loop) updates, while a meta-update (outer loop) optimizes this initialization across tasks.

\textbf{Test-time updates (key difference).} Unlike the other agents in this paper, MAML is allowed to \emph{learn on the test network}. When evaluated on an unseen topology, it first runs a small support set of episodes and performs a few gradient updates (inner-loop adaptation). We then measure performance on a separate query set using the adapted parameters.

\paragraph{State Representation}
To enable this adaptation in practice, we first transformed each state into a fixed-length 12-dimensional vector derived from the list of valid actions provided by the NetSecGame environment. Specifically, three features were derived from the source IP (the number of services running on the source host, whether the source belongs to a known network, and the centrality of the current source in the known graph). Six features were then extracted from the target IP (the attack progress on the target IP, whether the target is already under control, the number of services discovered on the target host, whether the host contains any known data, whether the target host is blocked from the source host, and the target's connectivity in the partial known topology). Finally, three global features were included to capture environment-level information available to the agent (the number of hosts currently under control, the number of hosts discovered, and the ratio of controlled hosts to discovered hosts), representing the agent's overall status.

These hand-designed features play a similar role to the \emph{concepts} used by the conceptual agent: they distill the raw observation into a small set of semantically meaningful signals (e.g., control status, progress stage, connectivity, and reachability constraints) that are intended to be less tied to specific IP values. While the conceptual agent enforces address invariance through an explicit abstraction mapping, the MAML agent relies on this feature representation and gradient-based adaptation to support transfer under unseen IP reassignment.

\paragraph{Policy Network}
Based on this state representation, we designed a policy network consisting of three fully connected layers. The input layer receives the 12-dimensional state vector (3 source IP features, 6 target IP features, and 3 global features), and maps it to 64 hidden neurons, followed by another hidden layer of 64 neurons, both using ReLU activation. The final output layer projects to five action types of the NetSecGame environment: ScanNetwork, FindServices, FindData, ExploitService, and ExfiltrateData.

For action selection, the policy produces a set of logits for the five action types associated with each candidate in the valid action list. For each candidate, we retain only the logit corresponding to its type, making one score per candidate. These scores are combined into a categorical probability distribution over all valid actions, from which the final action (i.e., the specific source-target candidate) is sampled.

\paragraph{Task Definition}

In the context of NetSecGame, we define a task for the MAML agent as a network topology with a fixed set of IP addresses and node structure. Within a task, each training episode randomizes both the starting node and the designated goal node, ensuring variability while preserving the overall network layout. This allows the agent to learn strategies that generalize across different initial conditions and objectives within the same topology. During meta-training, each task is further divided into a support set and a query set. The support set episodes are used for inner-loop adaptation, and the query set episodes are then used for outer-loop updates.

\paragraph{Training Procedure} 
We tried to align the training and testing settings with those of other agents as closely as possible. However, because of the distinct mechanism of the MAML, which relies on meta-training across multiple tasks, each consisting of separate support and query sets, the settings could not be matched exactly.

While other agents were trained on five tasks overall, the MAML agent uses five tasks for each epoch. For each task, the agent trains on 40 episodes in total; 30 support-set episodes are used for three inner-loop adaptation steps, followed by 10 query-set episodes used for the outer-loop update. 

\paragraph{Testing Procedure} 
The MAML agent was evaluated on a previously unseen task with a fixed topology but randomized start and goal nodes in Scenario 1. For each evaluation run, the agent first performed inner-loop adaptation on 50 support-set episodes per inner update step for three adaptation steps (i.e., 150 support episodes total), followed by evaluation on a query-set of 350 episodes using the adapted parameters. During testing, the outer loop didn't adapt the policy parameters; instead, only post-adaptation performance was measured after inner-loop adaptation to ensure a proper assessment of generalization. The primary evaluation metric was the win rate, calculated as the proportion of successful episodes in the query sets across unseen test tasks. 

\paragraph{Results}
MAML results are reported in Section~\ref{sec:results_adapt}.

\subsubsection{Reptile agent}
To provide a meta-learning baseline, we implemented a Reptile-based attacker agent. Reptile is a first-order gradient-based meta-learning algorithm that, similar to MAML, aims to learn a policy initialization that can be rapidly adapted to new tasks using a small number of task-specific updates \cite{nichol2018first}. Instead of using separate query-set gradients as in MAML, Reptile updates the shared initialization by moving it toward parameters obtained after task-level adaptation.

\textbf{Test-time updates.} Like MAML, Reptile is evaluated with inner-loop adaptation on the unseen topology: it uses a small support set of test episodes to update its parameters before being evaluated on held-out query episodes.

To isolate the effect of the meta-learning strategy itself, the Reptile agent uses the same state representation, policy network, action selection mechanism, and inner-loop adaptation procedure as the MAML agent. The main difference is in the outer loop update, which avoids second-order derivatives and is computationally cheaper.

\paragraph{Results}
Reptile results are reported in Section~\ref{sec:results_adapt}.

% (Moved) The detailed descriptions of the traditional learning agents (DQN/DDQN) are given earlier in this section under \textbf{Traditional learning agents}.

%-------------------------------------------
\section{Methodology}
\label{sec:methodology}

\subsection{Experimental setting and evaluation protocol}
\label{sec:method_protocol}
We evaluate generalization under a controlled distribution shift: the enterprise scenario, rewards, and action semantics are held fixed, while host and subnet IP addresses are reassigned. Agents are trained on five IP-range variants and evaluated on a sixth, held-out reassignment ($T_{unseen}$). We also evaluate on a seen configuration ($T_{seen}$) to separate generalization failure from pure learning failure.

At each step, the environment returns (i) an observation, including a reward, and whether the episode ended and why. Only the DDQN agent also asks, once at the beginning, for the complete set of valid parameterized actions in the whole game. The rest of the agents have the state and possible actions, so they can compute the possible parametrized possible actions on each state internally. Episodes terminate on success, failure, or a fixed horizon of 100 steps.

We report three primary aggregate metrics: win rate (success fraction), episodic return, and episode length (steps). Because these aggregates can hide qualitatively different failure modes, we additionally use behavioral signatures as a step-wise diagnostic (Section~\ref{sec:method_behavioral}).

Finally, because the compared agent families make different operational assumptions, Table~\ref{tab:comparability} summarizes their evaluation privileges and resource requirements. The \textbf{Cost} column is a qualitative proxy for inference-time burden, based on each agent's per-step computation: (i) small-network forward passes over hand-crafted features (Low), (ii) additional pretrained encoders and/or gradient-based test-time adaptation (Med/Med--High), and (iii) LLM calls per decision with long prompts and structured outputs (High). We do not claim a hardware-normalized cost comparison (e.g., GPU-seconds); instead, the table is used to interpret performance differences in light of model size, number of model invocations per step, and whether test-time learning is performed.

\subsection{Agent comparison dimensions}
\label{sec:method_agent_dimensions}
Different agent designs align with different offensive operating conditions. A (i) DQN-style learner is natural when an attacker can rehearse repeatedly on infrastructure that resembles the eventual target, e.g. cyber-ranges. (ii) Prompt-driven LLM-based approaches such as ReAct are appropriate when updating the model during the intrusion is undesirable, and logs from prior interactions are available for reasoning. Adaptive agents such as (iii) MAML or Reptile fit campaigns in which the attacker expects many related victims but only short dwell time in each. Knowledge gathered from earlier compromises provides an initialization that can be quickly adjusted once access to a new network is obtained. (iv) Concept-based agents are useful when the attacker can think in terms of machine roles rather than identifiers. Instead of remembering a specific IP address, the policy looks for whichever host functions as the authentication server or gateway, even when numbering or placement changes.

Table~\ref{tab:agent_differences} summarizes the main differences between agents in state/representation, action-space assumptions, training, and inductive biases.

\begin{table*}[!t]
\centering
\caption{Summary of agent differences in representation, training, action-space knowledge, and inductive biases.}
\small
\setlength{\tabcolsep}{4pt}
\begin{tabular}{
p{0.09\textwidth}
p{0.17\textwidth}
p{0.25\textwidth}
p{0.19\textwidth}
p{0.08\textwidth}
p{0.15\textwidth}
}
\toprule
\textbf{Agent} & \textbf{Key idea} & \textbf{State representation} & \textbf{Training} & \textbf{Knows $|\mathcal{A}|$?} & \textbf{Biases} \\
\midrule
Random & Random valid-action baseline & None; samples uniformly from valid actions & None & No & None \\
DQN & Value-based RL (no adaptation) & Candidate-action matrix with 12-D features from valid actions and partial graph & Fixed episodes & No & Feature design; action masking \\
MAML & Meta-learning for fast adaptation & Same candidate-action features as DQN & Meta-training (adaptation allowed) & No & Feature design; action masking \\
Reptile & First-order meta-learning & Same candidate-action features as DQN & Meta-training (adaptation allowed) & No & Feature design; action masking \\
DDQN & LM-based action scoring & JSON observation embedded by pretrained LM; Q-network scores state--action pairs & Train until convergence & Yes & Reward shaping; action masking \\
Conceptual & Role-based abstraction learning & Symbolic concept-state abstraction; Q-learning over concepts & Train until convergence & No & Concept abstraction; action pruning \\
ReAct & Prompted reasoning + acting & Text observation embedded in prompt (optional short memory) & None (prompt-only) & No & Prompt scaffold; action masking \\
LLM-BERT & Action classification + parameter filling & Prompt state; ModernBERT predicts action type and parameters & Offline fine-tuning from ReAct traces & No & Prompt scaffold; action masking \\
\bottomrule
\end{tabular}
\label{tab:agent_differences}
\end{table*}

\begin{table*}[!t]
\centering
\caption{Comparability table: evaluation privileges and resource assumptions across agent families. \textbf{Pretrain} indicates whether the agent relies on external pretrained models (beyond randomly initialized networks trained purely in NetSecGame). \textbf{TT learn} indicates whether parameters are updated on $T_{unseen}$ before evaluation. \textbf{Heuristics} summarizes hand-designed components (Feat.=feature engineering; Rules=rule-based abstractions/pruning; Rwd.=reward shaping; Prompt=prompt scaffolding/memory). \textbf{Cost} is a qualitative inference-time profile.}
\scriptsize
\setlength{\tabcolsep}{4pt}
\renewcommand{\arraystretch}{1.1}
\begin{tabular}{l c c c c c}
\toprule
\textbf{Agent} & \textbf{Pretrain} & \textbf{TT learn} & \textbf{Heuristics} & \textbf{Prune} & \textbf{Cost} \\
\midrule
Random     & --         & No  & --                 & No  & Low \\
DQN        & --         & No  & Feat.              & Yes & Low \\
DDQN       & Encoder    & No  & Rwd.               & No  & Med \\
Conceptual & --         & No  & Rules + Rwd.       & Yes & Low--Med \\
MAML       & --         & Yes & Feat.              & Yes & Med--High \\
Reptile    & --         & Yes & Feat.              & Yes & Med \\
ReAct      & LLM        & No  & Prompt             & No  & High \\
LLM-BERT   & LLM + BERT & No  & Prompt             & No  & High \\
\bottomrule
\end{tabular}
\label{tab:comparability}
\end{table*}

\subsubsection{State representation}
In our experiments, one of the main differences between agents is how they turn the NetSecGame observation into an internal state suitable for learning or decision making. The simplest baseline (random) does not construct any state representation at all: it ignores the observation and samples uniformly from the currently valid actions. At the other extreme, the DDQN agent consumes the raw observation as a serialized JSON string and relies on a pretrained language-model encoder to map the states and actions into dense vectors; a learned Q-network then scores each concatenated (state, action) embedding pair.

Between these extremes, the DQN and meta-learning agents (MAML and Reptile) use an explicit feature representation derived from the valid-action list and the partial knowledge graph. Rather than embedding the full JSON, they build a candidate-centric matrix where each valid (source, target) candidate is represented by a fixed 12-dimensional feature vector capturing source properties, target progress/visibility, and global discovery/ownership statistics. This representation is shared across DQN and the meta-learning policies, making the key distinction not the input encoding but whether the agent is allowed to adapt its parameters at test time (meta-learning) versus being evaluated as a fixed policy (DQN).

Finally, the conceptual agent represents state symbolically by abstracting away concrete identifiers such as IP addresses and network prefixes. It maps hosts, networks, services, and block relations into address-invariant ``concepts'' and performs Q-learning on this concept-state. 

LLM-based agents (ReAct and LLM-BERT) instead keep the state in textual form within prompts: ReAct directly reasons over the textual observation (optionally including a short rolling memory of recent actions/outcomes), while LLM-BERT retains the same prompt-level situational analysis but replaces generative action selection with fine-tuned ModernBERT modules for action-type classification and parameter filling.

\subsubsection{Action-space knowledge}
Another key distinction is whether an agent must assume a fixed, globally enumerated action space. Most agents operate solely on the \emph{currently valid} action list computed locally by the agent (Conceptual, ReAct, LLM-BERT, DQN, and the meta-learning variants), and therefore do not require advance knowledge of the total number of possible actions. In contrast, DDQN is implemented with a fixed action vocabulary and thus requires full knowledge of the complete action space in order to define its output dimensionality and Q-function parametrization.

\subsubsection{Training procedure}
The agents differ primarily in how (and whether) they are trained: DQN and the meta-learning approaches (MAML, Reptile) are trained for a fixed number of episodes, DDQN and the conceptual agent are trained until their learning converges, and the prompt-based LLM agent (ReAct) is not trained at all; the LLM-BERT variant also does not learn a policy online, but its ModernBERT components are fine-tuned offline using data collected from ReAct traces.

\subsubsection{Hand-designed biases}
Not all ``hand-designed'' components are heuristics in the same sense. We distinguish: (i) \emph{explicit heuristics/reward shaping} that directly modify the agent's objective or prune actions via fixed rules, and (ii) \emph{inductive biases} such as feature engineering, prompting structure, or action masking that constrain how decisions are represented rather than encoding domain rules.

Importantly, we treat these engineered components as \emph{part of the method} rather than as hidden advantages: in realistic cyber autonomy settings, agents are often deployed with scaffolding such as abstraction layers, action filters, interface constraints, and operator-provided priors. Our goal is therefore to compare complete agent designs (Table~\ref{tab:comparability}), while making the presence of heuristics explicit so that performance can be interpreted as a trade-off between robustness, transparency, and engineering effort.

The analysis of how each agent handles its biases is:
\begin{itemize}[noitemsep, leftmargin=*]
    \item \textbf{Random:} none.
    \item \textbf{Explicit heuristics / shaping:}
    \begin{itemize}[noitemsep, leftmargin=*]
        \item \textbf{DDQN:} intrinsic-reward shaping (bonus for newly discovered information) to mitigate sparse external rewards.
        \item \textbf{Conceptual:} rule-based concept abstraction (Tables~\ref{tab:concept-heuristics} and \ref{tab:action-heuristics}) and rule-based action pruning to remove redundant or structurally impossible actions (Table~\ref{tab:action-filters-short}).
    \end{itemize}
    \item \textbf{Inductive bias (not domain heuristics):}
    \begin{itemize}[noitemsep, leftmargin=*]
        \item \textbf{DQN / MAML / Reptile:} hand-crafted 12-D feature representation; additionally, these agents are effectively \emph{action-masked} because they only score the currently valid candidate actions.
        \item \textbf{ReAct / LLM-BERT:} prompt/interface scaffolding (textual state, structured prompting and/or short memory). In our setup they are also action-masked by restricting selection to the currently valid actions; this is an interface constraint rather than a cybersecurity-specific heuristic.
    \end{itemize}
\end{itemize}

\subsection{Behavioral signature comparison}
\label{sec:method_behavioral}
The evaluation of agent generalization in complex network environments requires a shift from aggregate performance metrics to granular behavioral diagnostics. We therefore extract and compare \emph{behavioral signatures}---the empirical distribution of action types over time---under the seen ($T_{seen}$) and unseen ($T_{unseen}$) IP assignments.

In the \textit{NetworkSecurityGame}, high-level metrics like cumulative reward often obscure the specific reasons for a policy's failure in novel environments. Success is contingent on a rigid hierarchy of prerequisite actions (e.g., scanning before exploitation). The goal of this comparison is to determine whether generalization failure stems from a total loss of policy logic or from a \emph{stalling} effect, where an agent persists in reconnaissance actions that are no longer effective under $T_{unseen}$.

The plots used to compare behavioral signatures in Section~\ref{sec:comparative_behavior} include a dotted line representing how many trajectories remain active at each step. Because a trajectory ends only when the game is won or the maximum episode horizon is reached (100 steps in our setup), this dotted line is also a proxy for the cumulative number of wins by step $t$. For example, in an experiment of 500 trajectories total, if the line has value 400 at step 20, it means that 100 trajectories terminated before step 20, which (under this termination rule) implies 100 wins before step 20.

\subsubsection{Data collection and processing}
To ensure the observed behaviors represent the learned policy rather than stochastic exploration, all evaluation episodes are collected with no random exploration ($\epsilon=0$). Data is gathered from two distinct sources:
\begin{itemize}[noitemsep, leftmargin=*]
    \item \textbf{Known environment ($T_{seen}$):} evaluation runs on a training task configuration to establish a behavioral baseline.
    \item \textbf{Unseen environment ($T_{unseen}$):} evaluation runs on the held-out task configuration to test generalization.
\end{itemize}

The collection process treats each agent as a fixed policy $\pi$ interacting with the environment transition dynamics $P(s'\mid s,a)$. Each episode (denoted $\tau \in T$) is a sequence of transitions $(s_0,a_0,s_1,a_1,\dots,s_T)$ sampled by following $\pi$ until a terminal state is reached or the maximum step limit is exceeded. Since $\epsilon=0$, variation in action frequencies is driven by differences in encountered states $s_t$ across episode seeds.

We aggregate episodes to compute the \emph{empirical distribution over action types} at each decision step $t$. We first map each executed \emph{parameterized} action (e.g., ``scan subnet 10.0.3.0/24'') to its corresponding \emph{action type} $a$ (e.g., \texttt{ScanNetwork}, \texttt{FindServices}, \texttt{FindData}, \texttt{ExploitService}, \texttt{ExfiltrateData}).

Let $T$ be the set of collected evaluation episodes and let $T_t \subseteq T$ denote the subset of episodes that are still active (not yet terminated) at step $t$. For a given action type $a$, we define
\begin{equation}
P(a_t \mid t, T) = \frac{1}{|T_t|} \sum_{\tau \in T_t} \mathbbm{1}\bigl(a_{t}^{\tau} = a\bigr),
\end{equation}
where $a_t^{\tau}$ is the action type chosen by the agent at step $t$ in episode $\tau$, and $\mathbbm{1}(\cdot)$ is an indicator that equals 1 if the chosen type matches $a$ and 0 otherwise. Thus, $P(a_t \mid t, T)$ is simply the fraction of still-running episodes whose step-$t$ action has type $a$; for each $t$, the probabilities over action types sum to 1.

\subsubsection{Visualization and interpretation}
We visualize each behavioral signature using a dual-axis temporal plot. The stacked bars show the action-type proportions over time, while the dashed line (secondary axis) shows a reachability metric: the fraction/number of trajectories that remain active at each step. By comparing the $T_{seen}$ and $T_{unseen}$ signatures, we localize the \emph{generalization gap}---the time step(s) where behavior diverges from the successful patterns observed during training.

\section{Results}
\label{sec:results}

This section reports performance on the seen and unseen IP reassignment variants and summarizes key failure modes across agent families. We organize results by agent type with shared learning and representation assumptions, using the common evaluation protocol and metrics defined in Section~\ref{sec:method_protocol}. A consolidated cross-scenario comparison table is provided in Section~\ref{sec:overall_comparison}.

\subsection{Traditional learning agents}
\label{sec:results_traditional}

\subsubsection{DQN agent}
We evaluate both the single-replay and dual-replay DQN variants described in Section~\ref{subsec:DQN_agent} under the two standard evaluation regimes defined in our protocol (Section~\ref{sec:method_protocol}): (i) transfer under IP reassignment (train on five IP variants, test on a held-out reassignment) and (ii) a same-topology sanity check (train and test on the same configuration).

% Table~\ref{tab:summary_dqn} reports the average training vs. evaluation win rate across these configurations, highlighting the sharp drop under IP reassignment compared to same-topology evaluation. Though the dual buffer agent outperforms the single buffer agent in both configurations, both agents' evaluation win rates drop by an average of $28.03 \%$. This change in win rate shows the struggle of the agent to generalize well to an unseen topology, even after removing the IP from the state embedding. 

% We tested the agent's ability to learn and generalize by allowing it to explore more and start exploiting at a lower rate (epsilon). This change did not shift the evaluation win rate for any of the configurations. In the dual buffer we found that increasing the success ratio (the number of successful instances taken from the success buffer) allowed the win rate to be consistently higher by an average of $1.8\%$ , not much but at least there was a change. Ultimately, DQN was a good baseline to compare the MAML agent that uses the same state embedding. 

Table~\ref{tab:summary_dqn} summarizes the average training and evaluation win rates across configurations. While the dual-buffer agent consistently outperforms the single-buffer variant, both agents experience a substantial decline under IP reassignment, with evaluation win rates dropping by an average of $28.03\%$ relative to same-topology evaluation. This degradation indicates that the agents struggle to generalize to unseen network topologies, even when IP information is removed from the state embedding. Overall, the DQN agent serves as a useful baseline for comparison with the MAML-based agent, which uses the same state representation.

% We further evaluated whether increasing exploration, by extending the exploration phase and delaying exploitation (lower $\epsilon$), would improve generalization. However, this modification produced no measurable improvement in evaluation performance. In the dual-buffer setting, increasing the success ratio (the proportion of samples drawn from the success buffer) yielded a modest but consistent gain, improving win rates by an average of $1.1\%$. Overall, the DQN agent serves as a useful baseline for comparison with the MAML-based agent, which uses the same state representation but is explicitly designed to improve cross-topology generalization.

\begin{table}[h]
\centering
\footnotesize
\setlength{\tabcolsep}{6pt}
\caption{Average Win Rate for DQN Experiments}
\label{tab:summary_dqn}
\resizebox{\columnwidth}{!}{%
\begin{tabular}{ccc}
  \toprule
  \textbf{Exp \#} &
  \shortstack{\textbf{Final Training WR}} &
  \shortstack{\textbf{Final Evaluation WR}} \\
  \midrule
  1 & $37.18 \pm 0.91$ & $2.07 \pm 2.35$ \\
  2 & $47.60 \pm 4.98$ & $25.20 \pm 9.81$ \\
  3 & $34.55 \pm 2.78$ & $2.47 \pm 3.07$ \\
  4 & $51.53 \pm 5.50$ & $35.40 \pm 1.73$ \\
  \bottomrule
\end{tabular}%
}
\end{table}

\subsubsection{DDQN agent}
The DDQN agent achieved a 100\% win rate in all five training environments but failed to generalize, reaching a 0\% win rate in the unseen test environment. We attribute this gap primarily to the state embedding: observations that are semantically identical but differ only in IP addresses appear far apart in embedding space.

\begin{figure}[h]
    \centering
    \includegraphics[width=0.9\linewidth]{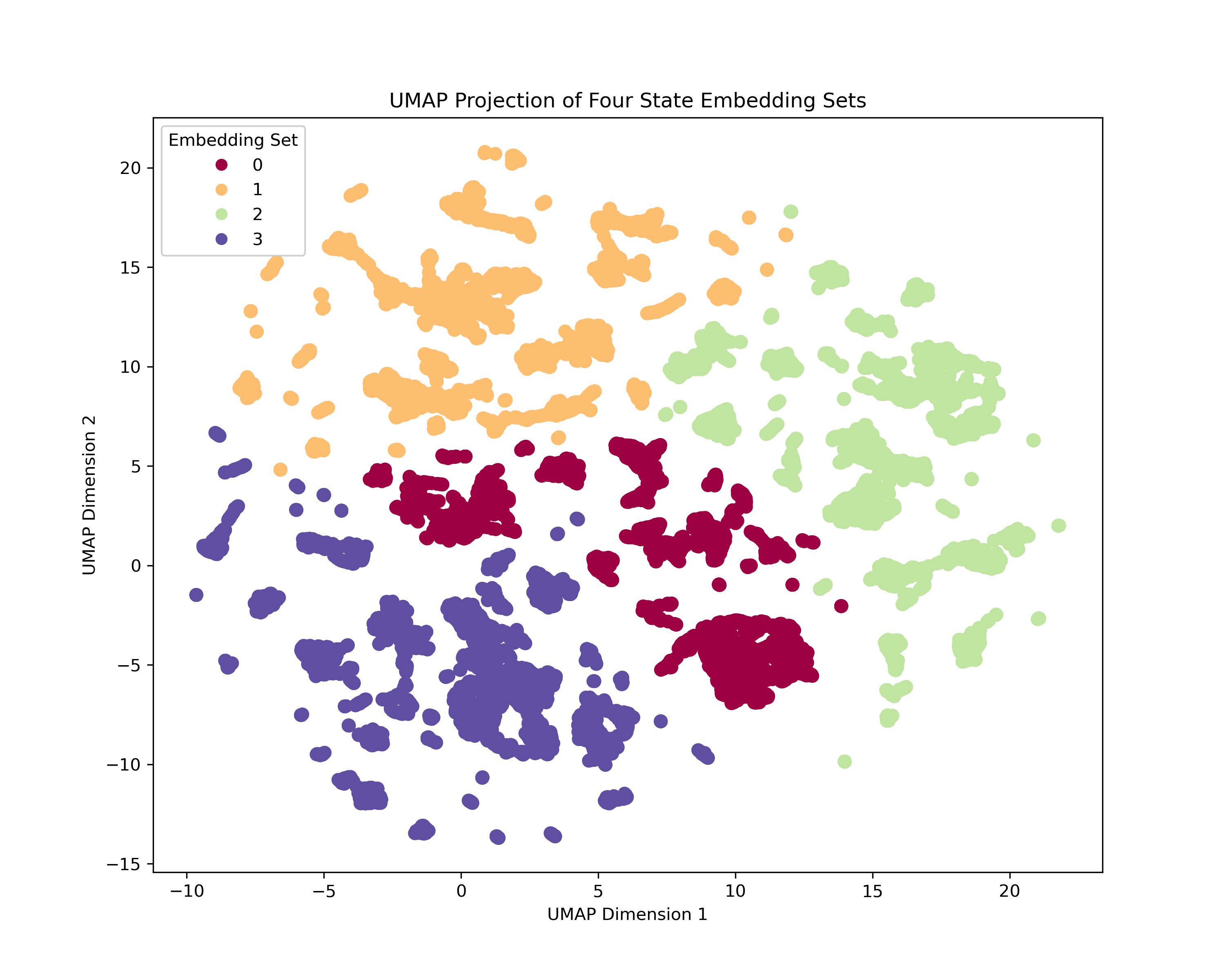}
    \caption{UMAP projection of state embeddings for the same scenario under four different IP ranges.}
    \label{fig:umap_states}
\end{figure}

To illustrate this effect, we embedded states from the same topology under four different IP ranges and projected them with UMAP. Figure~\ref{fig:umap_states} shows four clearly separated clusters, indicating that the encoder does not normalize away IP-specific tokens; consequently, the agent cannot reliably treat two states with different IPs as equivalent.

\vspace{-15pt}

\subsection{LLM-based agents}
\label{sec:results_llm}

\subsubsection{ReAct agent}
\begin{figure}[H]
    \centering
    \includegraphics[width=0.9\columnwidth]{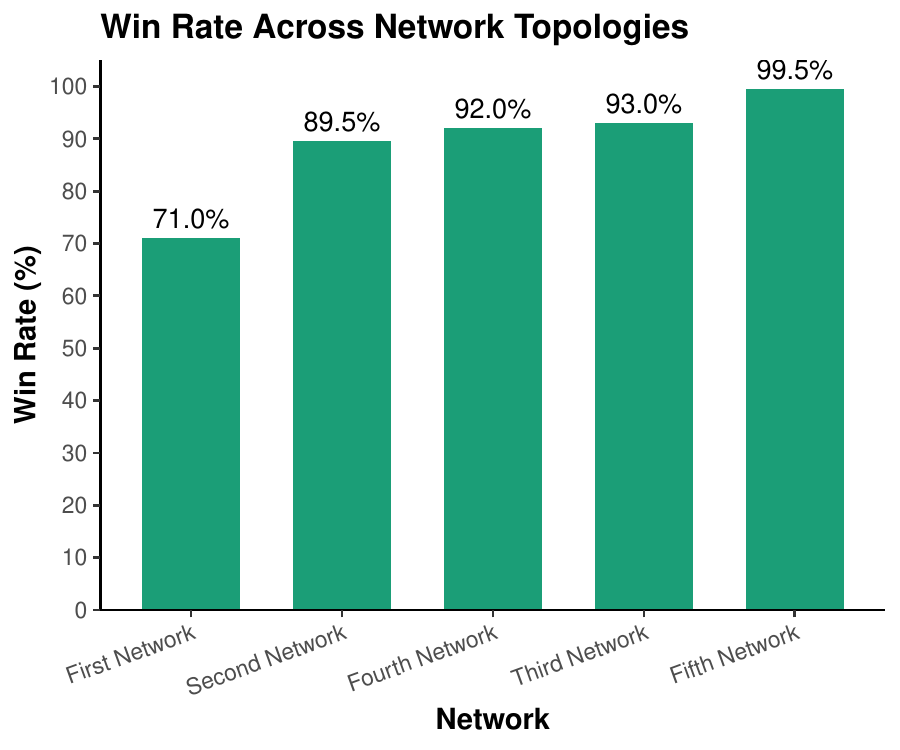}
    \caption{Win rate performance of the ReAct agent.}
    \label{fig:react-network-performance}
\end{figure}

For the ReAct agent, Table~\ref{tab:summary_react} summarizes the key performance metrics across all episodes, as well as separately for winning and losing episodes. 
ReAct agent performance exhibited substantial variability across all five training network scenarios, as seen in Figure~\ref{fig:react-network-performance}.

\begin{table}[h]
\centering
\caption{Episode Summary Statistics ReAct Agent using GPT-OSS-120b on the Unseen Topology}
\label{tab:summary_react}
\footnotesize
\setlength{\tabcolsep}{3pt}
\resizebox{\columnwidth}{!}{%
\begin{tabular}{@{}lrrr@{}}
  \toprule
  \textbf{Metric} & \textbf{All Episodes} & \textbf{Wins} & \textbf{Losses} \\
  \midrule
  Count & 263 & 250 & 13 \\
  Win Rate (\%) & 95.1 & --- & --- \\
  Avg Steps & $31.2 \pm 23.2$ & $27.8 \pm 18.2$ & $96.7 \pm 2.4$ \\
  Avg Total Reward & $+63.9 \pm 40.9$ & $+72.2 \pm 18.2$ & $-97.5 \pm 4.3$ \\
  Min Steps & 8 & 8 & 91 \\
  Max Steps & 100 & 91 & 100 \\
  \bottomrule
 \end{tabular}%
}
\end{table}
\vspace{-5pt}

The agent achieved a win rate of 95.1\%, completing 250 out of 263 episodes successfully. Winning episodes achieved an average total reward of $+72.2 \pm 18.2$, while losing episodes accumulated an average penalty of -97.5, resulting in an overall average reward of $+63.9 \pm 40.9$ across all episodes. The step count ranges also reveal that successful completions occurred with an average of $27.8 \pm 18.2$ steps, while failure episodes reached $96.7 \pm 2.4$.

\subsubsection{LLM-BERT agent}
Table~\ref{tab:bert_seen_vs_unseen} summarizes the key performance metrics on the previously unseen network topology. The agent achieved a win rate of 51.6\%, completing 129 out of 250 episodes successfully. We also compare against the metrics in seen topologies to show that the generalization gap is minimal.

\begin{table}[h]
\centering
\caption{LLM-BERT Performance Comparison: Seen vs. Unseen Topology}
\label{tab:bert_seen_vs_unseen}
\footnotesize
\setlength{\tabcolsep}{4pt}

\resizebox{0.8\columnwidth}{!}{%
\begin{tabular}{@{}lcc@{}}
\toprule
\textbf{Metric} & \textbf{Seen (T5)} & \textbf{Unseen} \\
\midrule
\multicolumn{3}{c}{\textit{Overall}} \\
\midrule
Win Rate (\%) & 52.2 & 51.6 \\
Return & $+12.5\pm81.6$ & $-6.5\pm81.7$ \\
Steps & $39.7\pm36.7$ & $57.7\pm33.6$ \\
\midrule
\multicolumn{3}{c}{\textit{Wins}} \\
\midrule
Return & $+87.1\pm6.8$ & $+70.8\pm14.3$ \\
Steps & $12.9\pm6.8$ & $29.2\pm14.3$ \\
\midrule
\multicolumn{3}{c}{\textit{Losses}} \\
\midrule
Return & $-69.0\pm33.4$ & $-89.0\pm18.6$ \\
Steps & $69.0\pm33.4$ & $88.0\pm17.8$ \\
\bottomrule
\end{tabular}%
}
\end{table}

\subsubsection{Analysis of LLM Agents}
\begin{figure}[h]
\centering

\begin{subfigure}{\columnwidth}
    \centering
    \includegraphics[width=0.8\columnwidth]{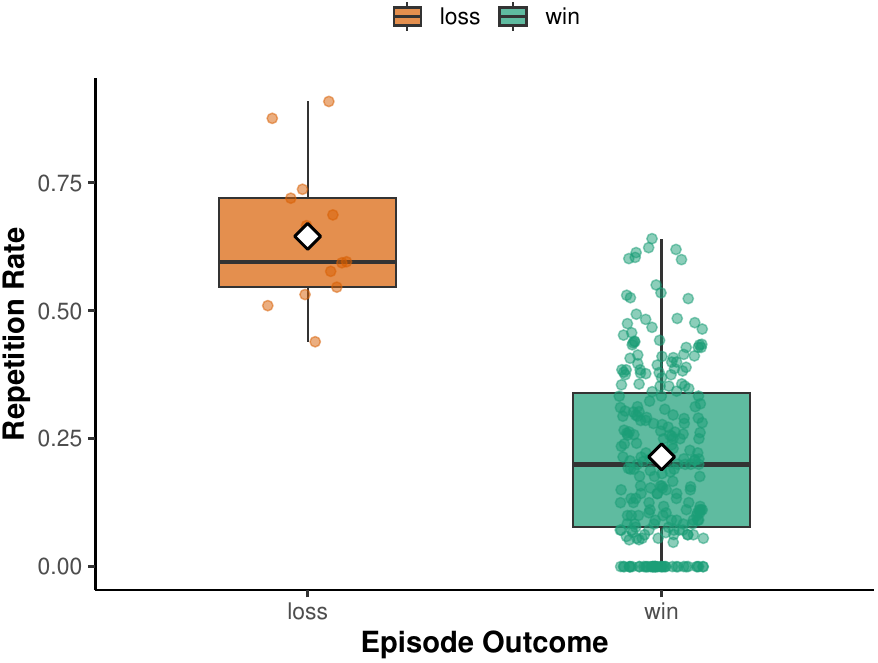}
    \caption{ReAct agent repetition behavior.}
    \label{fig:react-repetition_analysis}
\end{subfigure}
\vspace{-0.5mm}

\begin{subfigure}{\columnwidth}
    \centering
    \includegraphics[width=0.8\columnwidth]{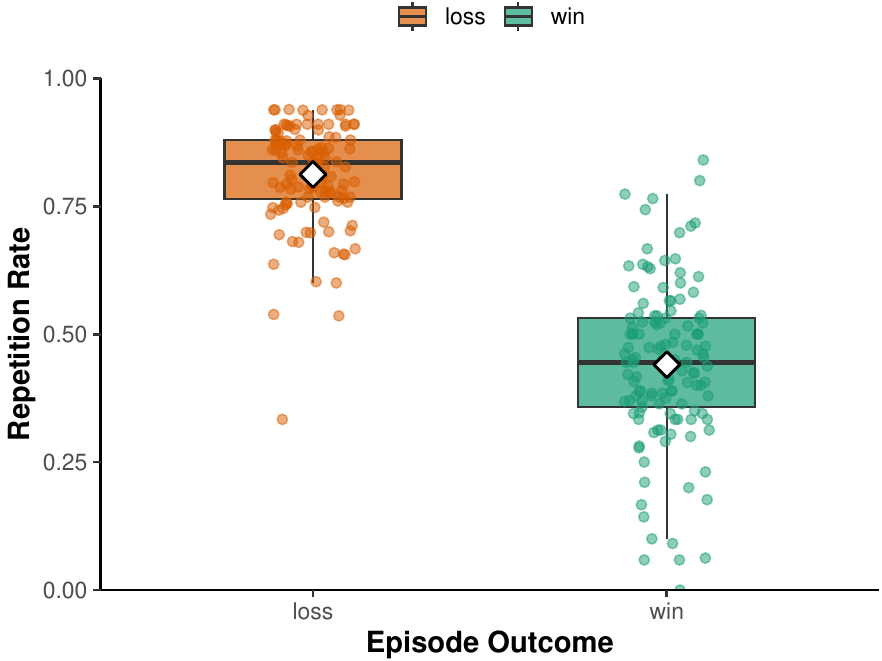}
    \caption{LLM-BERT agent repetition behavior.}
    \label{fig:bert-repetition_analysis}
\end{subfigure}

\vspace{-2mm}
\caption{Repetition behavior analysis: distribution of repetition rates across successful and failed episodes for the ReAct and LLM-BERT agents.}
\label{fig:repetition_analysis}

\end{figure}

\paragraph{ReAct Agent:} The dominant cause of failure was the exhaustion of the internal limit actions, accounting for 92.3\% (12 out of 13) of all losses. These episodes averaged 96.4 steps and -96.42 rewards, indicating that the agent became stuck attempting invalid JSON actions, as seen in Table \ref{tab:loss_types}. The remaining failure (1 episode) resulted from reaching the external step limit. This was mainly due to a higher number of actions being repeated in episodes that resulted in losses, as shown in Figure~\ref{fig:react-repetition_analysis}.

\begin{table}[h]
\centering
\caption{Loss Type Analysis for ReAct using GPT-OSS-120b on the Unseen Network Topology}
\label{tab:loss_types}
\footnotesize
\setlength{\tabcolsep}{5pt}

\resizebox{\columnwidth}{!}{%
\begin{tabular}{@{}lrrrr@{}}
\toprule
\textbf{Loss Type} &
\shortstack{\textbf{Count}} &
\shortstack{\textbf{\% of}\\\textbf{Losses}} &
\shortstack{\textbf{Avg}\\\textbf{Steps}} &
\shortstack{\textbf{Avg}\\\textbf{Reward}} \\
\midrule
Invalid Actions Exhausted & 12 & 92.3 & 96.4 & -96.42 \\
Step Limit Reached & 1 & 7.7 & 100.0 & -110.00 \\
\midrule
\textbf{Total} & \textbf{13} & \textbf{100.0} & \textbf{96.8} & \textbf{-97.54} \\
\bottomrule
\end{tabular}%
}
\end{table}

\begin{figure*}
    \centering
    \includegraphics[width=0.9\textwidth]{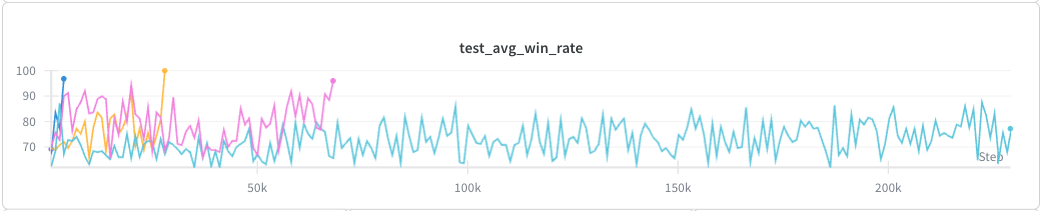}
    \caption{Conceptual Agent: Performance of each of the five training runs. Runs 0, 1, 2, and 4 stopped after reaching the threshold of 0.95 evaluation win rate. Run number 3 stopped after reaching 230,000 episodes without reaching the 0.95 evaluation win rate.}
    \label{fig:conceptual_agent_eval_perf}
\end{figure*}

\begin{figure*}
    \centering
    \includegraphics[width=0.9\textwidth]{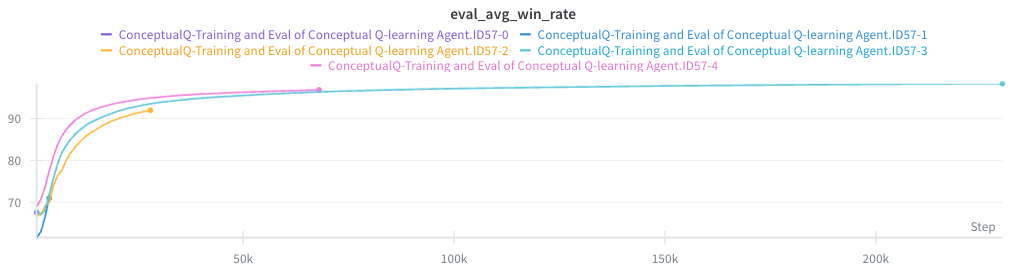}
    \caption{Conceptual Agent: Performance of each of the five training runs during training episodes. The win rate is computed while the model is still learning and changing.}
    \label{fig:conceptual_agent_train_perf}
\end{figure*}

\paragraph{LLM-BERT agent:} Most failures of the LLM-BERT agent occur due to negative returns (90.9\%) rather than reaching the step limit, indicating that the agent often takes suboptimal actions that accumulate penalties as seen in Table \ref{tab:bert_loss_types}. These episodes are also longer on average (88 steps), suggesting difficulty recovering once the agent deviates from an effective strategy. Repetition analysis (Figure \ref{fig:bert-repetition_analysis}) further shows that loss episodes have substantially higher action repetition rates than successful ones, and repetition increases with episode length, indicating that many failures arise from repeatedly executing similar actions without making meaningful progress.

\begin{table}[h]
\centering
\caption{Loss Type Analysis for LLM-BERT on Unseen Network Topology}
\label{tab:bert_loss_types}
\footnotesize
\setlength{\tabcolsep}{4pt}
\begin{tabular}{@{}lrrr@{}}
\toprule
\textbf{Loss Type} & \textbf{Count (\%)} & \textbf{Steps} & \textbf{Reward} \\
\midrule
Negative Return & 110 (90.9) & 86.5 & -89.93 \\
Step Limit & 11 (9.1) & 100.0 & -85.25 \\
\midrule
\textbf{Total} & \textbf{121 (100)} & \textbf{88.0} & \textbf{-89.25} \\
\bottomrule
\end{tabular}
\end{table}

\subsection{Generalization and adaptation agents}
\label{sec:results_adapt}

\subsubsection{Conceptual agent}
\paragraph{Training}
We report two notions of win rate during learning. Figure~\ref{fig:conceptual_agent_eval_perf} shows the \emph{evaluation} win rate for each training run, computed by periodically \emph{freezing} the current Q-table/policy and running evaluation episodes with learning disabled; this isolates policy quality from on-policy exploration noise. In contrast, Figure~\ref{fig:conceptual_agent_train_perf} reports the win rate measured \emph{during} training episodes while the agent is still updating its Q-values.

\paragraph{Testing:} Table~\ref{tab:conceptual-testing-results} reports the \emph{test} win rate of the final, \emph{frozen} conceptual policy on the held-out unseen topology (scenario 6), repeated over 100 random seeds. The conceptual agent achieved an average win rate of 65.53\% (std.~4.77\%), indicating that its policy generalizes reliably to the held-out scenario rather than depending on favorable initialization. In addition to the success rate, the agent produced strongly positive average returns (620.83 on average), but with substantial variability across seeds (average return std.~519.88), which suggests that while the agent typically finds effective strategies, the quality of those strategies can vary noticeably from run to run. Episodes required 67.16 steps on average (std.~30.26), and successful episodes were shorter at 49.89 steps on average (std.~23.21), implying that when the agent wins it tends to do so more efficiently than in the average rollout. Overall, the conceptual agent provides the strongest adaptation performance among the generalization-focused baselines in our experiments, combining a high win rate with comparatively efficient successful trajectories.

\begin{table}[h]
\centering
\caption{Conceptual agent performance in unseen scenario 6 over 100 seeds.}
\label{tab:conceptual-testing-results}
\footnotesize
\setlength{\tabcolsep}{6pt}

\resizebox{0.8\columnwidth}{!}{%
\begin{tabular}{lcc}
\toprule
\textbf{Metric} & \textbf{Average} & \textbf{Std. Dev.} \\
\midrule
Win Rate (\%) & 65.530 & 4.770 \\
Returns & 620.830 & 519.884 \\
Episode Steps & 67.158 & 30.260 \\
Win Steps & 49.894 & 23.209 \\
\bottomrule
\end{tabular}%
}

\end{table}

\begin{figure}[h]
\centering

\begin{subfigure}{\columnwidth}
    \centering
    \includegraphics[width=\linewidth]{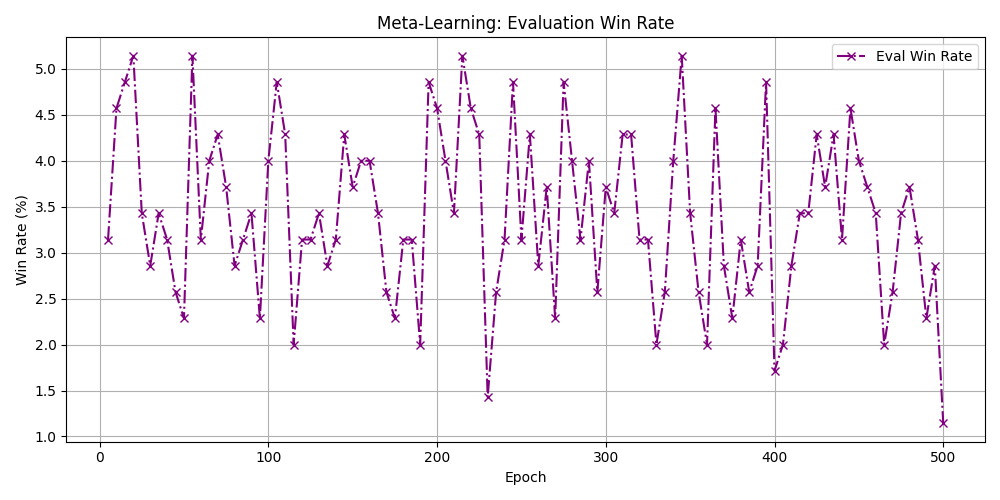}
    \caption{Reptile agent win rate}
    \label{fig:reptile_winrate}
\end{subfigure}

\begin{subfigure}{\columnwidth}
    \centering
    \includegraphics[width=\linewidth]{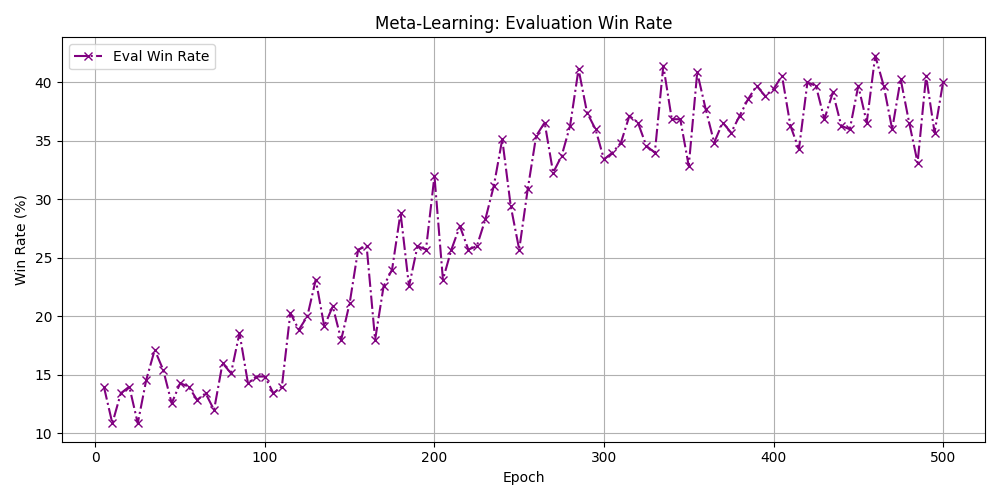}
    \caption{MAML agent win rate}
    \label{fig:maml_winrate}
\end{subfigure}
\caption{Training win rate comparison of meta-learning agents.}
\label{fig:meta_winrate}
\end{figure}

\subsubsection{Reptile agent}
We trained the Reptile agent as a baseline for meta-adaptive agents. The Reptile agent did not exhibit stable learning behavior or consistent performance gains across tasks. As shown by its win-rate trajectory (Figure~\ref{fig:reptile_winrate}), training exhibited large oscillations without a sustained upward trend, indicating that the learned initialization did not reliably support fast adaptation in this setting. Consequently, Reptile was not competitive with the conceptual or MAML agents in terms of generalization performance.

\subsubsection{MAML agent}
\paragraph{Results:} We trained the MAML agent across 500 epochs (a total of 100,000 episodes) to analyze its meta-learning performance in the NetSecGame environment. The meta-loss steadily decreased from around 60 to the low 45s over 500 epochs, indicating successful optimization of the policy. The evaluation win rate increased from roughly 10--15\% at the start to 35--42\% by the end of training as seen in Figure \ref{fig:meta_winrate}. Overall, MAML learns a transferable initialization that improves with training, but it remains substantially less effective than the conceptual agent on the unseen scenario, suggesting that additional task diversity and/or adaptation steps may be needed to close the gap.

\begin{table*}[h]
\centering
\footnotesize
\renewcommand{\arraystretch}{1.15}

\caption{Comparison of different generalization approaches and baseline agents in the testing (previously unseen) topology. The episode timeout is 100 steps, and the maximum possible return is 95.}

\begin{tabular}{lccccc}
\toprule
\textbf{Agent} &
\shortstack{\textbf{Win}\\\textbf{rate (\%)}} &
\shortstack{\textbf{Episode}\\\textbf{return}} &
\shortstack{\textbf{Episode}\\\textbf{steps}} &
\shortstack{\textbf{Total}\\\textbf{training episodes}} &
\shortstack{\textbf{\# of steps}\\\textbf{for win episode}} \\
\midrule

Random agent (baseline)  & $6.0 \pm 2.4$ & $-100.93 \pm 34.95$ & 98.47 & -- & -- \\
Single Buffer DQN   &$2.07 \pm2.36 $  & $-106.71\pm3.69$ & $ 98.98\pm1.11 $ & 5000 & $10.33 \pm1.79 $ \\
Dual Buffer DQN  &$3.07\pm2.72$& $-104.45\pm5.01$     & $97.82\pm2.02 $ & 5000 & $ 15.33\pm 13.59 $ \\
DDQN+emb & 0.0 & -99.00 & 100.00 & -- & -- \\

\midrule
ReAct (gpt-oss-120b) & $95 \pm 22$ & $63.86 \pm 40.85$ & $31.16 \pm 23.21$ & -- & $27.76 \pm 18.20$ \\
LLM-BERT & $51.60 \pm 3.16$ & $-6.50 \pm 81.70$ & $57.66 \pm 33.57$ & -- & $29.16 \pm 14.29$ \\

\midrule
Reptile (baseline) & $2.76 \pm 1.47$ & $-105.90 \pm 2.06$ & $98.94 \pm 0.46$ & 100,000 & $58.42 \pm 10.7$ \\
MAML & $40.00 \pm 0.86$ & $-50.80 \pm 1.21$ & $84.80 \pm 0.28$ & 100,000 & $61.99 \pm 0.24$ \\
Conceptual Q-learning & $65.53 \pm 4.7$ & $62.0 \pm 51.9$ & $67.1 \pm 30.2$ & 331,000 & $49.8 \pm 23.2$ \\

\bottomrule
\end{tabular}

\label{tab:gen_results}
\end{table*}

\subsection{Overall comparison of Results}
\label{sec:overall_comparison}
This subsection summarizes overall performance and generalization outcomes on the previously unseen topology. Table~\ref{tab:gen_results} compares success rate, efficiency (steps), and achieved return across all agent families.
\paragraph{Baselines and traditional RL} The random agent provides a reference point for how hard the task is without a learned strategy: it succeeds only $6.0\%$ of the time and runs essentially to the 100-step timeout on average (98.47 steps), producing strongly negative return. The value-based baselines fail to transfer under IP reassignment: DDQN+emb achieves a $0\%$ win rate and hits the 100-step limit with near-minimum return, consistent with a policy that does not recover a viable attack plan when identifiers change. (Q-learning and DQN results are not reported in Table~\ref{tab:gen_results}.)

\paragraph{LLM-based agents} ReAct is the strongest performer on the unseen topology, with a reported win rate of $95\%$ and short episodes (31.16 steps on average; 27.76 steps for wins). Its positive mean return (63.86) indicates that most episodes reach the goal without incurring large failure penalties. However, the large win-rate variance ($\pm 22$) suggests sensitivity to stochasticity (e.g., decoding, initial conditions, or action-validity edge cases), motivating the failure analyses in Section~\ref{sec:results_llm}. LLM-BERT provides a more resource-efficient alternative to pure LLM action generation, but its overall return is close to zero and slightly negative ($-6.50$) despite a moderate win rate (51.60\%). This gap between win rate and mean return indicates that failures are typically long and costly (timeouts or extended unproductive loops), which is consistent with the repetition/invalid-action behaviors diagnosed earlier.

\paragraph{Generalization and adaptation agents} Among the learning-based approaches, conceptual Q-learning is the strongest: it reaches a $65.53\%$ win rate with a positive mean return (62.0) on the unseen topology, demonstrating that explicit address-invariant abstraction can preserve a usable attack policy across reassignment. The trade-off is efficiency and training cost: episodes are longer (67.1 steps on average; 49.8 steps in wins) than ReAct, and training required substantially more experience (331k episodes). MAML achieves a lower win rate (40\%), but with lower number of observations, and remains inefficient (84.8 mean steps; 62.0 steps in wins) with a negative mean return, suggesting that the learned initialization and limited test-time adaptation do not consistently discover a high-quality strategy under this shift. However, MAML observes a particular training scenario much less than Conceptual Agents. Reptile performs near the random baseline (2.76\% win rate, near-timeout episode lengths, and very negative return), indicating that first-order meta-learning was not sufficient in this environment and training regime.

\paragraph{Takeaway} Table~\ref{tab:gen_results} shows a clear ordering on the unseen topology: (i) prompt-based LLM reasoning (ReAct) yields the highest success and the most efficient winning trajectories, (ii) conceptual abstraction is the strongest among learning-based generalization mechanisms but requires substantially more training experience and still produces longer rollouts, and (iii) meta-learning (MAML/Reptile) provides at best partial transfer here, with MAML being more reliable, while traditional value-learning baselines do not generalize under identifier reassignment.

The results also suggest a clear deployment tradeoff across agent types. (i) If no prior knowledge of the target network or topology is available, LLM-based agents such as ReAct provide the most practical starting point since they require no environment-specific training and can operate in previously unseen settings. However, their behavior can be unstable and failure cases often involve repetitive or inefficient actions. (ii) When a topology similar to the deployment environment is available, conceptual adaptive agents offer the most reliable option, producing stable and interpretable behavior after environment-specific training. (iii) If the exact deployment topology is unknown but multiple related environments are available for training, meta-learning approaches such as MAML provide a middle ground by enabling rapid adaptation to new scenarios with only a few updates. 

In operational security settings, the choice therefore depends primarily on the level of prior environmental knowledge: zero knowledge favors LLM agents, topology-specific knowledge favors conceptual agents, and limited but some knowledge across multiple environments favors meta-learning approaches.

\section{Comparative Behavioral Analysis}
\label{sec:comparative_behavior}

This section compares results using the behavioral-signature methodology introduced in Section~\ref{sec:method_behavioral}. We group agents by family and contrast how their action-selection dynamics change from the seen to the unseen IP reassignment.

\subsection{Traditional learning agents}
\label{sec:behav_traditional}
We first compare agents that learn value functions or policies without any explicit mechanism for address invariance. For each agent, we analyze the seen vs. unseen pair of behavioral-signature plots and explain what changes under IP reassignment.

\subsubsection{Random baseline}
\label{sec:behav_random}
\begin{figure}[!t]
\centering
\begin{subfigure}[t]{\columnwidth}
\centering
\includegraphics[width=\columnwidth,height=0.18\textheight,keepaspectratio]{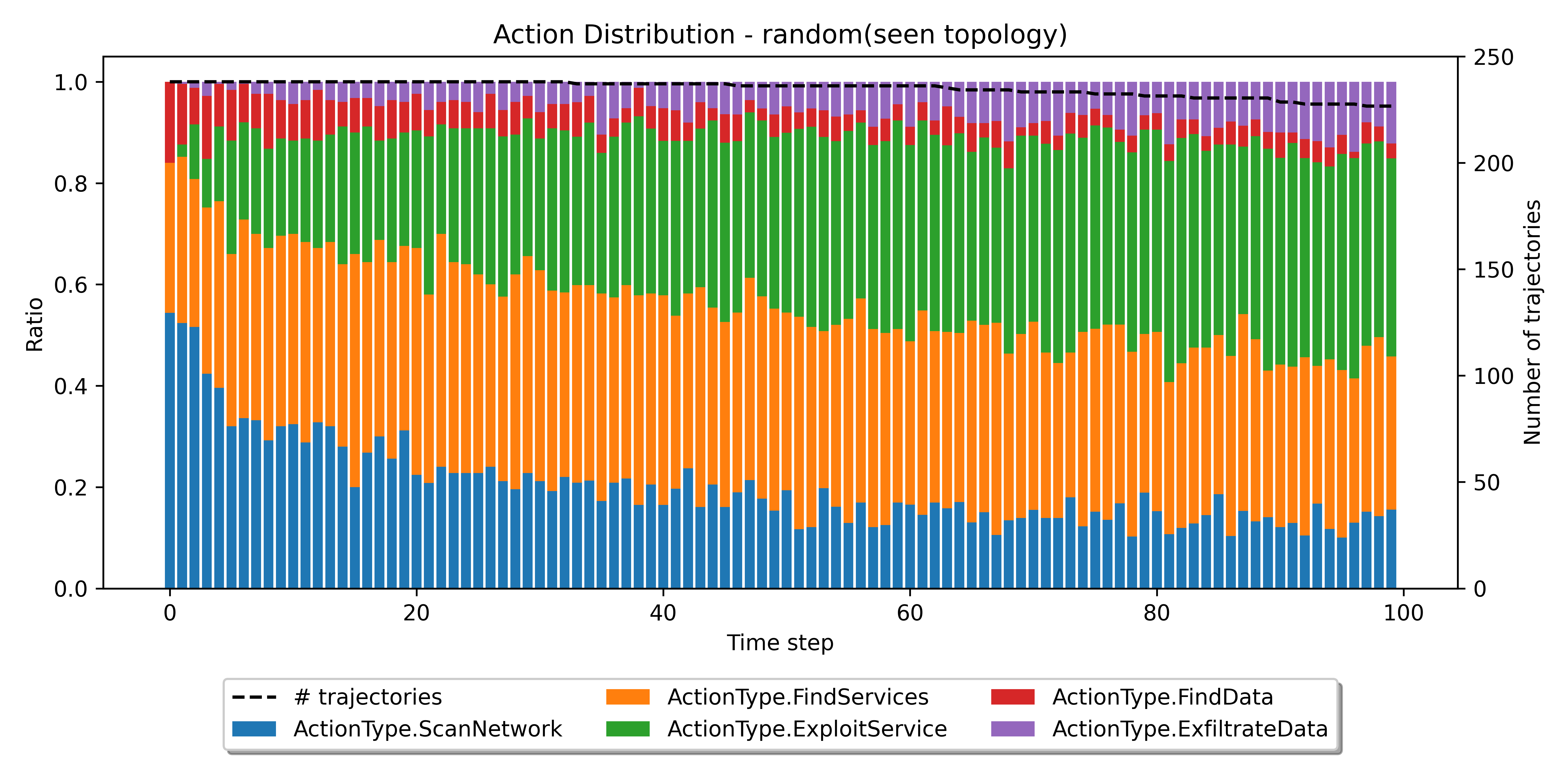}
\caption{Known topology ($T_{seen}$)}
\label{fig:rand_seen}
\end{subfigure}
\vspace{0.5ex}
\begin{subfigure}[t]{\columnwidth}
\centering
\includegraphics[width=\columnwidth,height=0.18\textheight,keepaspectratio]{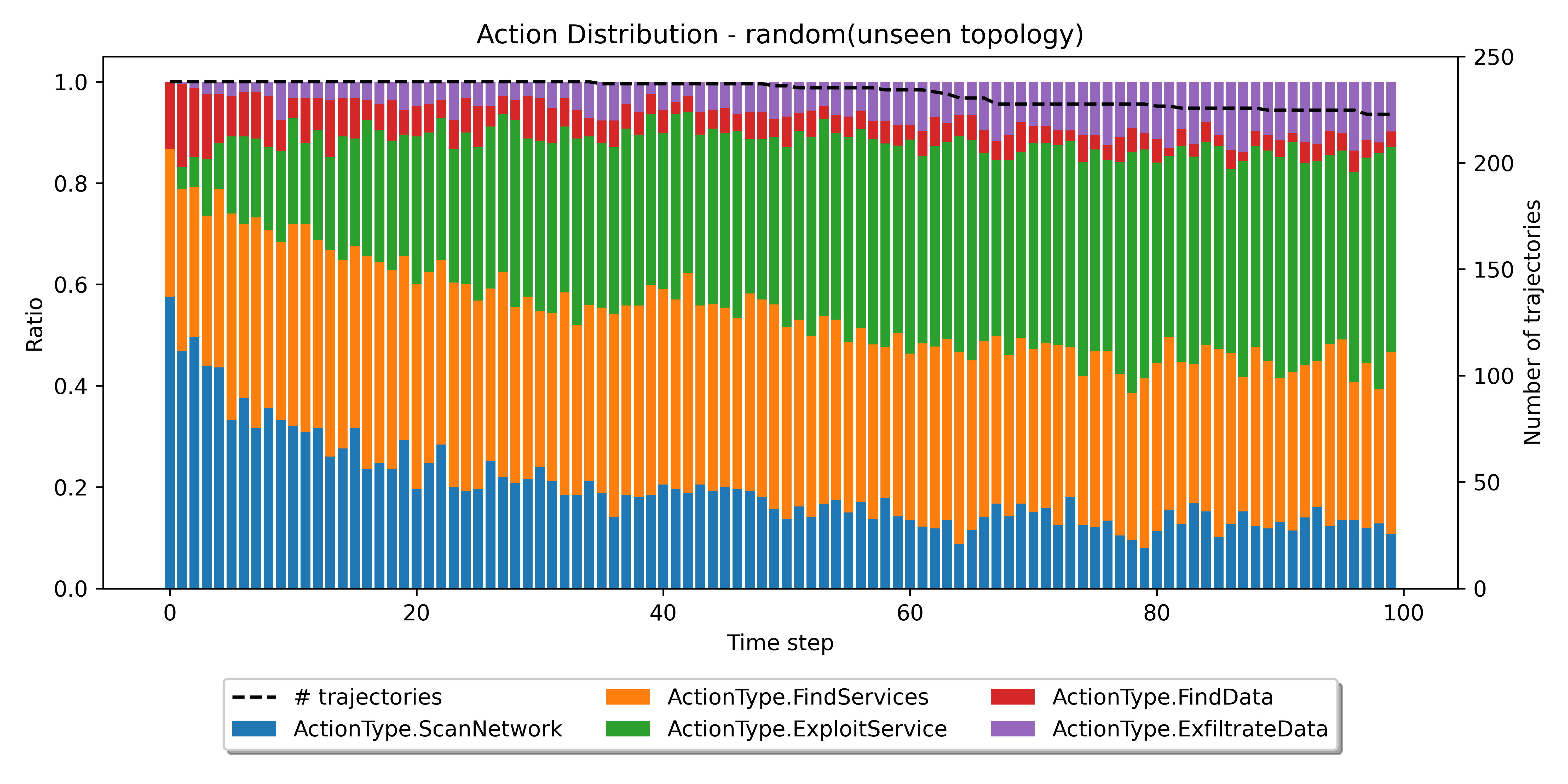}
\caption{Unseen topology ($T_{unseen}$)}
\label{fig:rand_unseen}
\end{subfigure}
\caption{Random baseline: action-type distribution over time on seen vs. unseen topology.}
\label{fig:behav_random_pair}
\end{figure}

Figures~\ref{fig:rand_seen}--\ref{fig:rand_unseen} show that even a random policy does not produce a uniform action distribution over time. This is an \emph{environment effect}: the valid-action set is constrained by prerequisites, so action types like \textit{ExploitService} and especially \textit{ExfiltrateData} are structurally unavailable until the agent has first discovered hosts, services, and data. As a result, the early steps are dominated by scanning/discovery actions and only later steps contain a mixture that includes higher-impact actions.

Crucially, the seen and unseen plots look similar because the random agent has no internal representation that can overfit to concrete IP identifiers. This makes it a useful reference signature: any systematic seen-to-unseen divergence for learned agents is attributable to the policy/model rather than to changes in action availability.

\subsubsection{Single Buffer DQN agent}
\label{sec:behav_dqn}
\begin{figure}[!t]
\centering
\begin{subfigure}[t]{\columnwidth}
\centering
\includegraphics[width=\columnwidth,height=0.18\textheight,keepaspectratio]{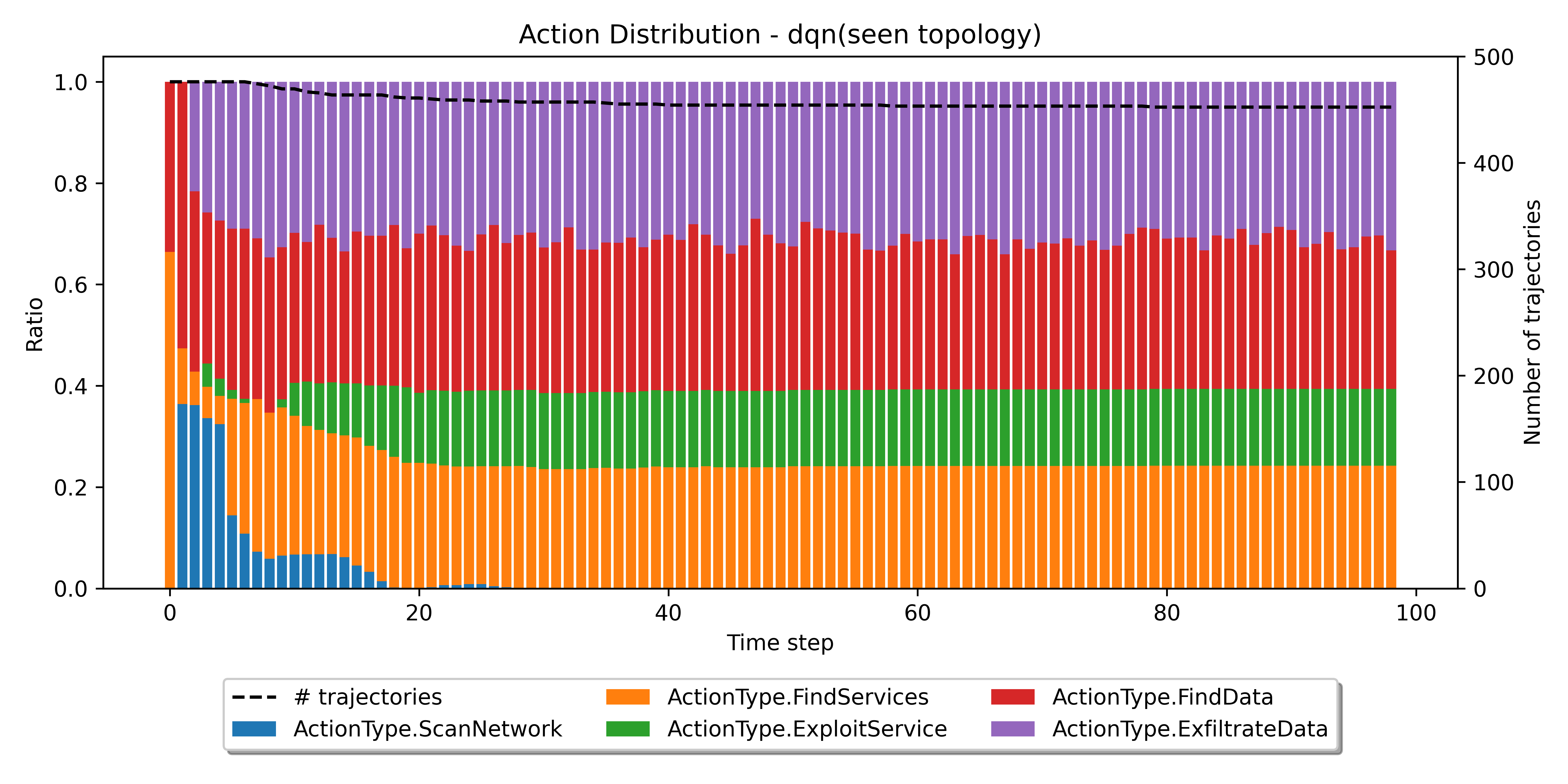}
\caption{Known topology ($T_{seen}$)}
\label{fig:dqn_seen}
\end{subfigure}
\vspace{0.5ex}
\begin{subfigure}[t]{\columnwidth}
\centering
\includegraphics[width=\columnwidth,height=0.18\textheight,keepaspectratio]{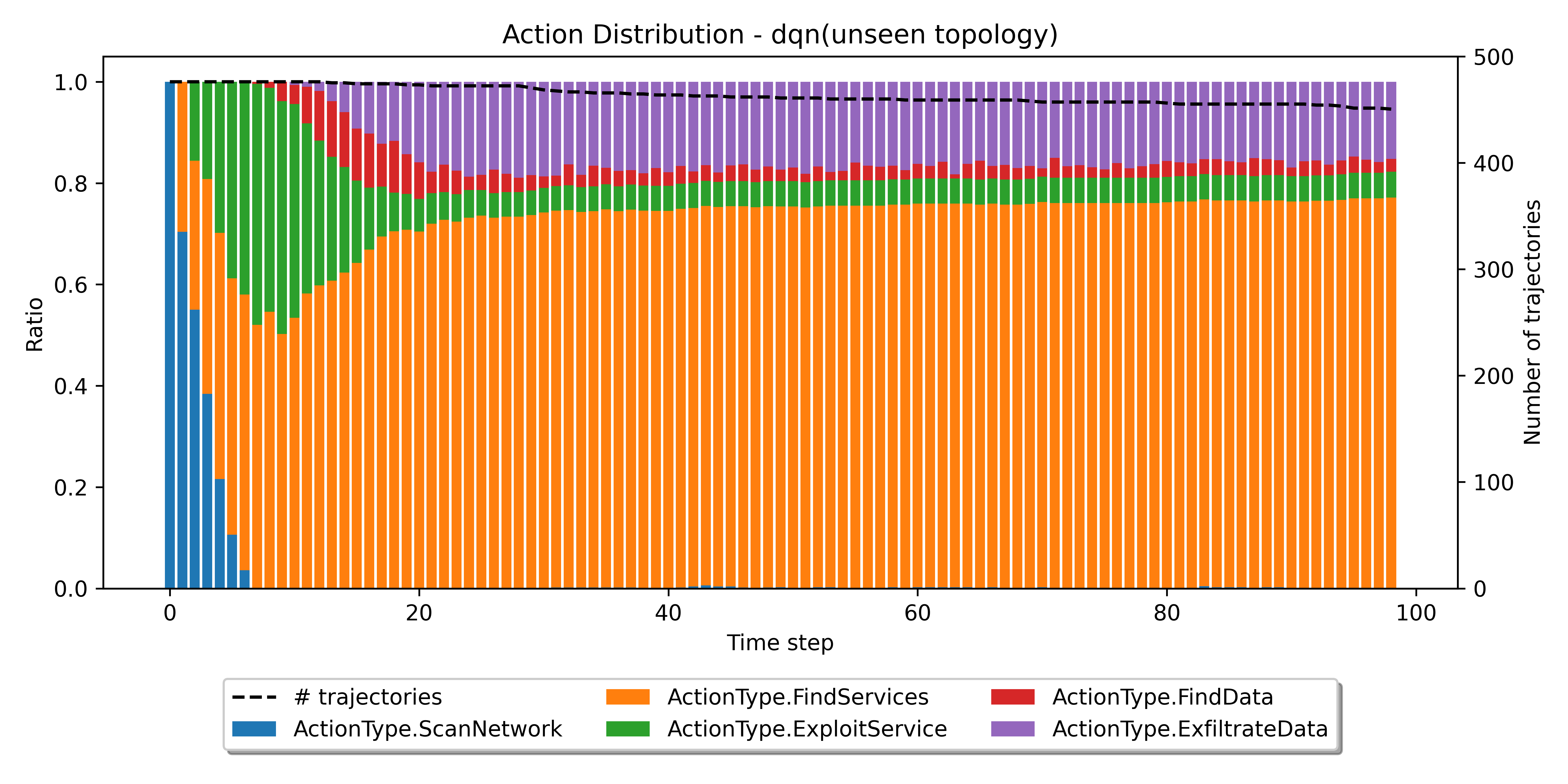}
\caption{Unseen topology ($T_{unseen}$)}
\label{fig:dqn_unseen}
\end{subfigure}
\caption{DQN: action-type distribution over time on seen vs. unseen topology.}
\label{fig:behav_dqn_pair}
\end{figure}

On $T_{seen}$ (Figure~\ref{fig:dqn_seen}), the signle buffer DQN signature shows a recognizable phase transition: the early horizon is dominated by discovery actions, after which the policy increasingly allocates probability mass to exploitation and downstream goal actions. This is consistent with the agent learning a long-horizon plan structure (scan $\rightarrow$ exploit $\rightarrow$ exfiltrate) on the training configuration. However, the agent failed to learn pass the step number 18, after which the probabilities of the actions are fixed. The win rate of this agent is lower than the random agent, suggesting that the learning was not only not happening but actually impaired.

On $T_{unseen}$ (Figure~\ref{fig:dqn_unseen}), this phase structure collapses. The distribution remains skewed toward early reconnaissance actions for substantially longer, and many trajectories persist until near the timeout (reachability remains high at late steps).
The original learned policy to start with FindServices and FindData was replaced with ScanNetwork, which is better in the unseen topology, but was curiously absent in the first steps of the seen topology. The FindService action (exploration) dominated the unseen topology actions,  matching the failure mode expected under IP reassignment: when state signatures change, the learned Q-function cannot reliably generalize and the greedy policy is effectively driven by poorly calibrated values, leading to repeated low-impact discovery actions rather than progressing toward exploitation and exfiltration. Despite the difference in the application of the policy, the win rate in the unseen topology remained similar to the training one, showing that the bias in the policy were larger than the differences intrudoced by the change in IP reassignment.

\subsubsection{DDQN agent}
\label{sec:behav_ddqn}
\begin{figure}[!t]
\centering
\begin{subfigure}[t]{\columnwidth}
\centering
\includegraphics[width=\columnwidth,height=0.18\textheight,keepaspectratio]{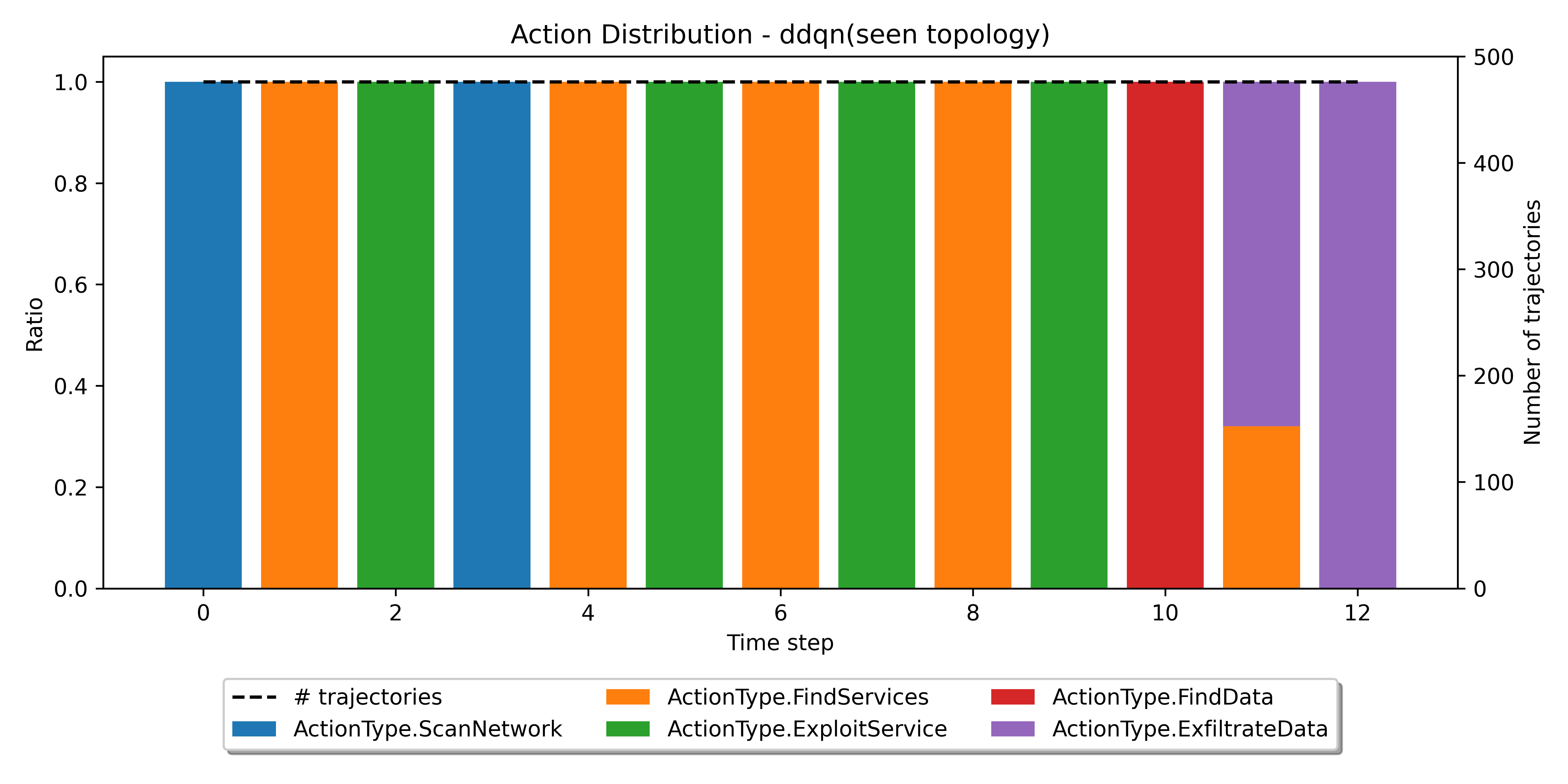}
\caption{Known topology ($T_{seen}$)}
\label{fig:ddqn_seen}
\end{subfigure}
\vspace{0.5ex}
\begin{subfigure}[t]{\columnwidth}
\centering
\includegraphics[width=\columnwidth,height=0.18\textheight,keepaspectratio]{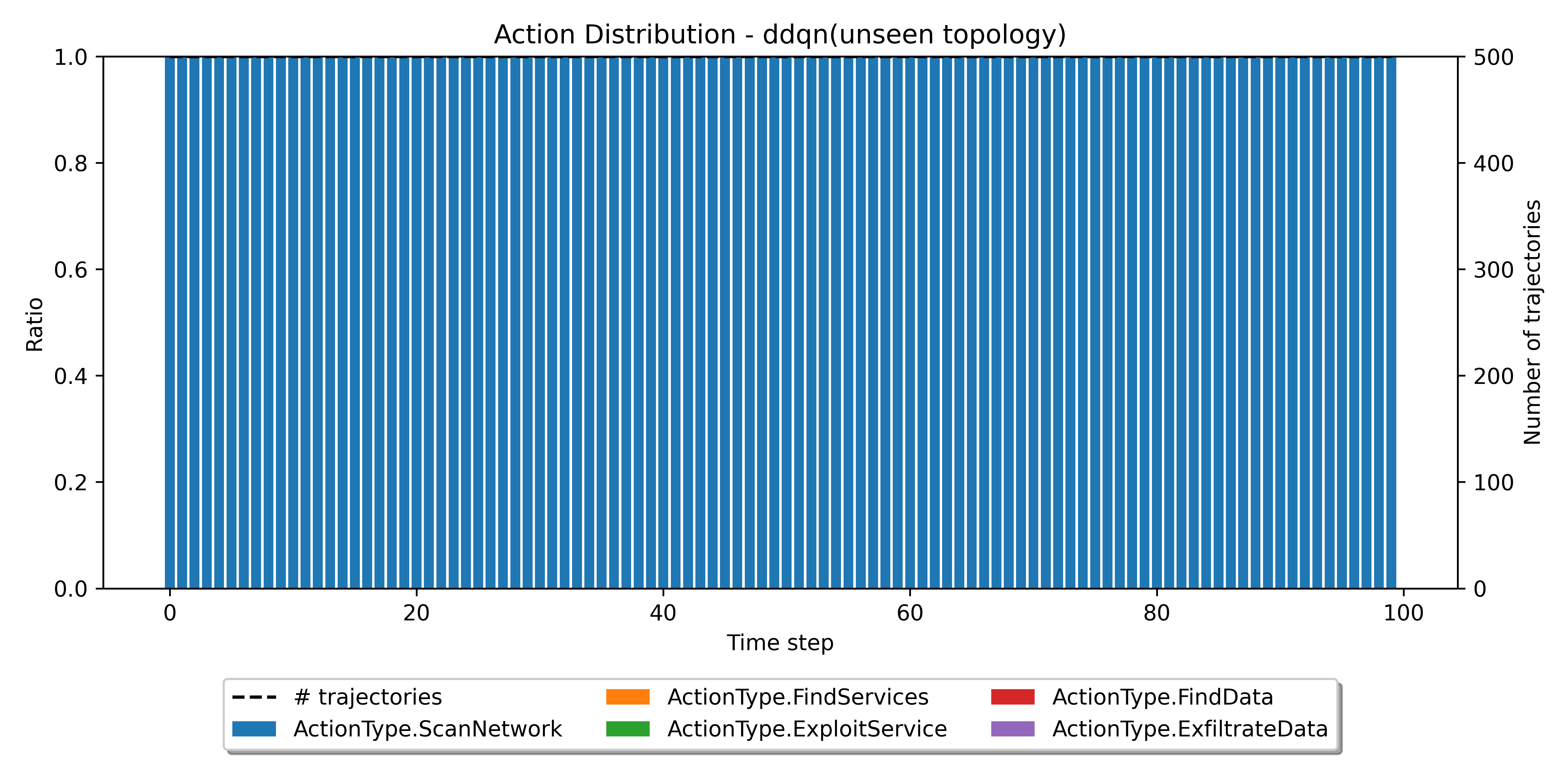}
\caption{Unseen topology ($T_{unseen}$)}
\label{fig:ddqn_unseen}
\end{subfigure}
\caption{DDQN: action-type distribution over time on seen vs. unseen topology.}
\label{fig:behav_ddqn_pair}
\end{figure}

DDQN exhibits the strongest seen-to-unseen divergence. On $T_{seen}$ (Figure~\ref{fig:ddqn_seen}), the agent slowly transitions from scanning to exploitation in later-stage actions, indicating that it can follow a somehow coherent plan when evaluated under the same identifier distribution it was trained on. However, its win rate in $T_{seen}$ was zero.

On $T_{unseen}$ (Figure~\ref{fig:ddqn_unseen}), the signature becomes fully dominated by early-phase actions with no evidence of sustained exploitation/exfiltration phases until near the end. This is consistent with the representation-level overfitting highlighted by the embedding analysis of the UMAP clusters in Section~\ref{sec:results}: if the state encoder separates semantically equivalent states by IP tokens, then the downstream Q-network receives out-of-distribution embeddings and its action preferences degrade into unproductive loops.

\subsection{LLM-based agents}
\label{sec:behav_llm}
We next analyze agents that reason over a textual description of the state using LLMs and select actions from the valid-action list on each state.

\subsubsection{ReAct}
\label{sec:behav_react}
\begin{figure}[!t]
\centering
\begin{subfigure}[t]{\columnwidth}
\centering
\includegraphics[width=\columnwidth,height=0.18\textheight,keepaspectratio]{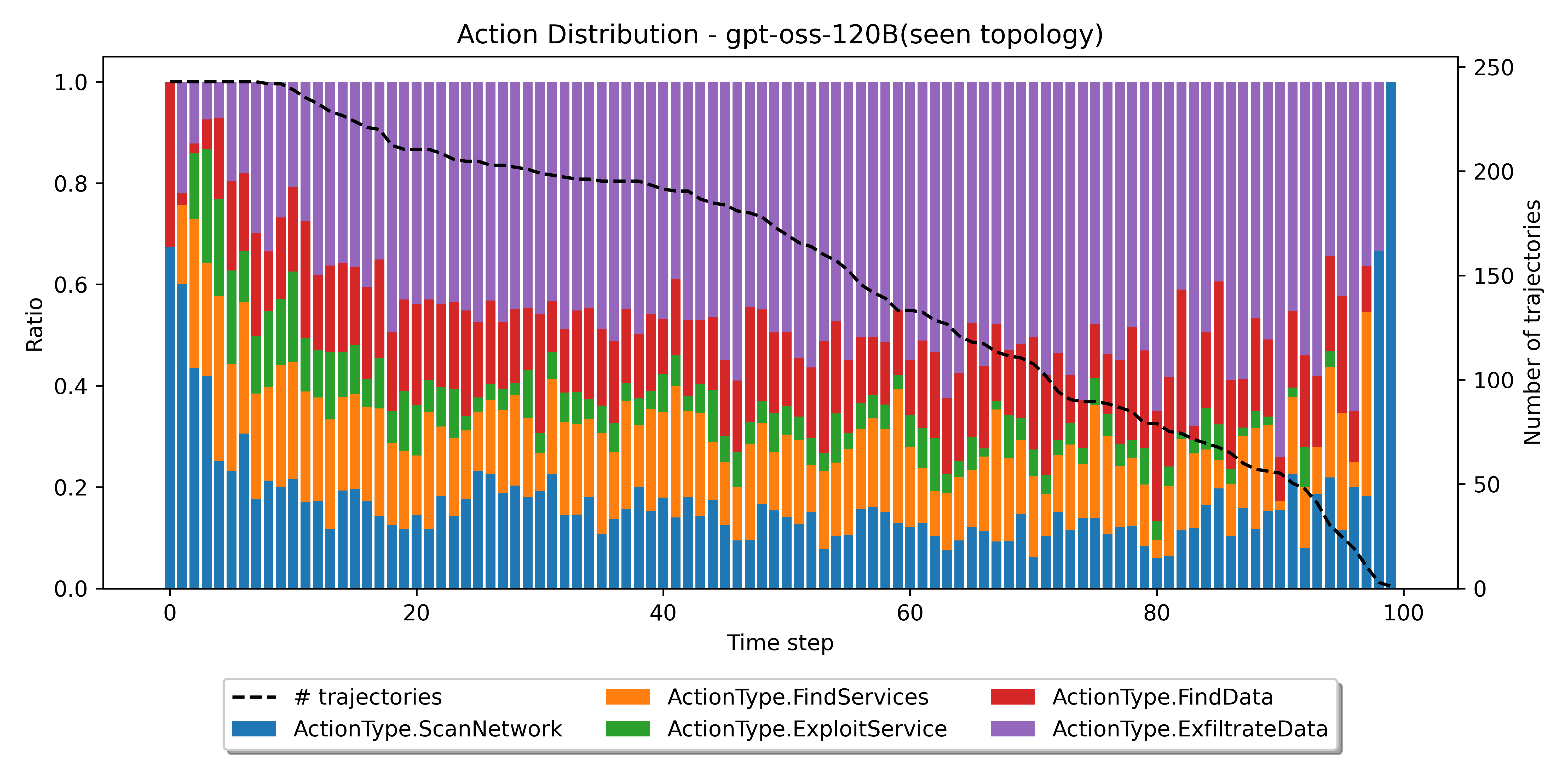}
\caption{Known topology ($T_{seen}$)}
\label{fig:react_seen}
\end{subfigure}
\vspace{0.5ex}
\begin{subfigure}[t]{\columnwidth}
\centering
\includegraphics[width=\columnwidth,height=0.18\textheight,keepaspectratio]{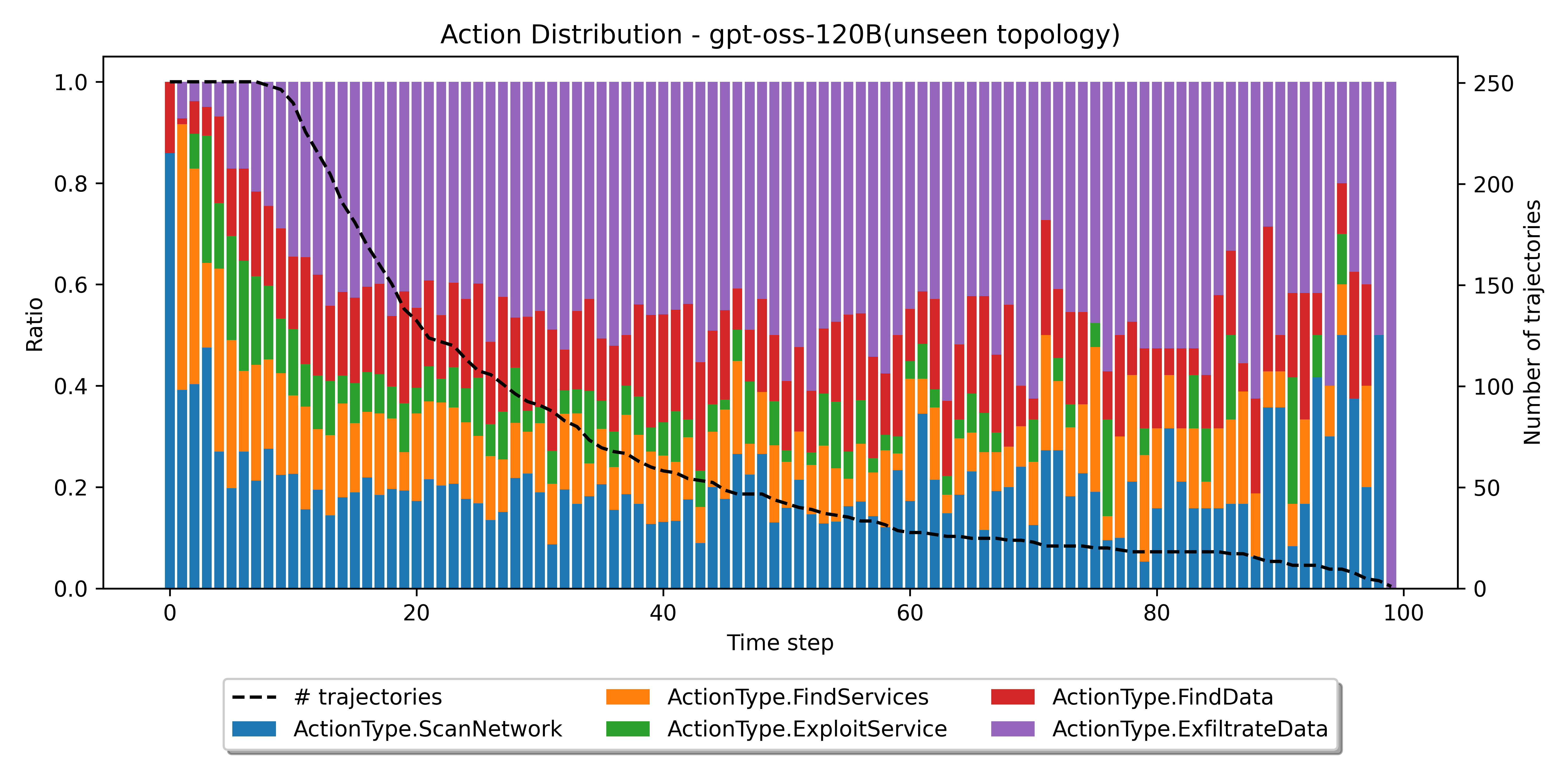}
\caption{Unseen topology ($T_{unseen}$)}
\label{fig:react_unseen}
\end{subfigure}
\caption{ReAct: action-type distribution over time on seen vs. unseen topology.}
\label{fig:behav_react_pair}
\end{figure}

ReAct shows comparatively stable phase structure across $T_{seen}$ and $T_{unseen}$ (Figures~\ref{fig:react_seen}--\ref{fig:react_unseen}). The early steps are dominated by scanning/discovery, after which the policy allocates substantial mass to exploitation and goal actions. This aligns with its high win rate: the model can re-derive a plausible next step from the current textual observation even when concrete identifiers are permuted. The main differences between $T_{seen}$ and $T_{unseen}$ are that the win rate during $T_{seen}$ is more gradual and linear while the win rate during $T_{unseen}$ is more sudden after 10 steps.

A likely explanation for this difference is architectural: ReAct is a prompt-driven policy that selects from a \emph{textual} list of currently valid actions using short-horizon context (the prompt window of recent actions and outcomes). Under IP reassignment, semantically equivalent choices are expressed with different surface-form identifiers and can appear in different positions/orderings in the valid-action list. These presentation changes can shift which actions the model treats as salient early in the episode, yielding either (i) faster completion of prerequisites and an earlier cascade into exploit/exfiltrate actions (a sharper early drop in reachability), or (ii) more exploratory or redundant reconnaissance before committing (a slower, more linear drop). Thus, the reachability-curve shape difference is consistent with prompt-level sensitivity to action-candidate presentation and short-term memory effects, rather than a change in the underlying task logic.

The remaining failures are possibly not primarily a \emph{planning} failure but an \emph{execution/interface} failure: once the agent enters a bad loop (e.g., repeated invalid or redundant actions), the reachability curve stays high and trajectories run close to the horizon. This matches the failure analyses reported for ReAct in the Results section.

\subsubsection{LLM-BERT}
\label{sec:behav_llmbert}
\begin{figure}[!t]
\centering
\begin{subfigure}[t]{\columnwidth}
\centering
\includegraphics[width=\columnwidth,height=0.18\textheight,keepaspectratio]{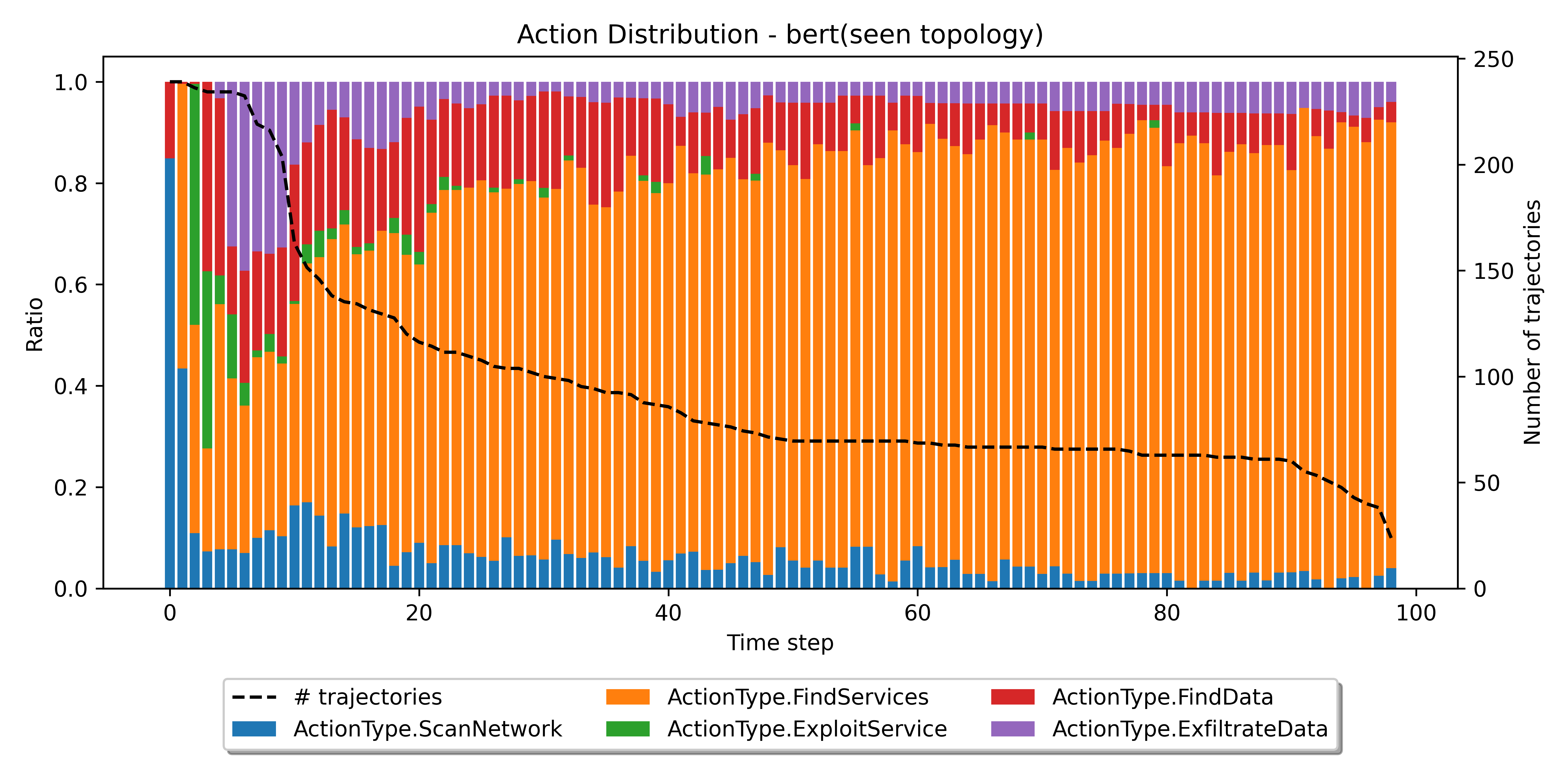}
\caption{Known topology ($T_{seen}$)}
\label{fig:llmbert_seen}
\end{subfigure}
\vspace{0.5ex}
\begin{subfigure}[t]{\columnwidth}
\centering
\includegraphics[width=\columnwidth,height=0.18\textheight,keepaspectratio]{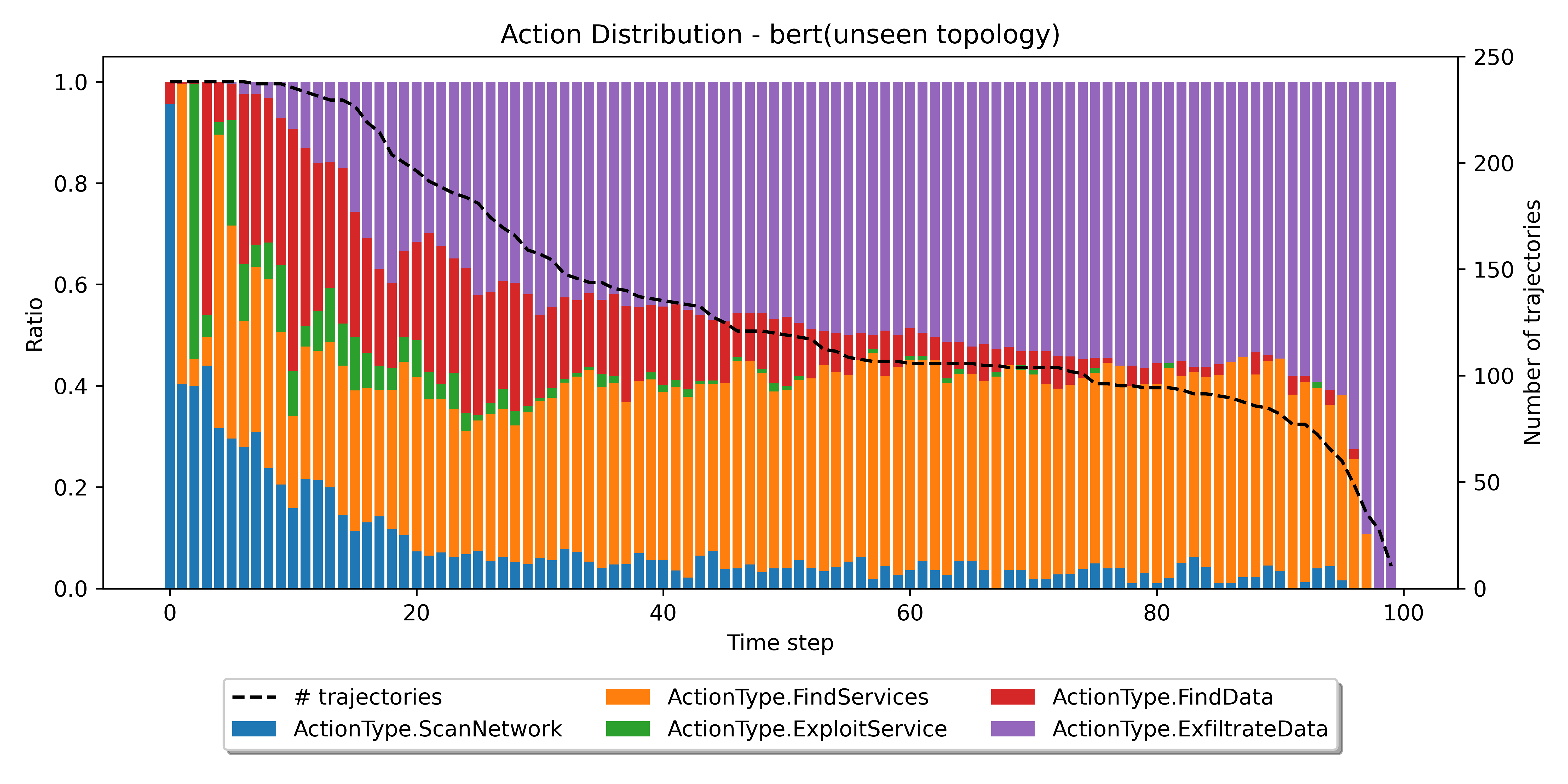}
\caption{Unseen topology ($T_{unseen}$)}
\label{fig:llmbert_unseen}
\end{subfigure}
\caption{LLM-BERT: action-type distribution over time on seen vs. unseen topology.}
\label{fig:behav_llmbert_pair}
\end{figure}

LLM-BERT retains part of ReAct's phase structure, but its behavior differs sharply between $T_{seen}$ and $T_{unseen}$ (Figures~\ref{fig:llmbert_seen}--\ref{fig:llmbert_unseen}). On $T_{seen}$ (Figure~\ref{fig:llmbert_seen}), the signature is dominated by \texttt{FindServices} for most of the horizon, indicating that the agent often fails to transition reliably into later-stage exploitation/exfiltration actions. In contrast, on $T_{unseen}$ (Figure~\ref{fig:llmbert_unseen}) the policy allocates substantially more mass to \texttt{ExfiltrateData}, suggesting that it more frequently completes the prerequisite chain and reaches the goal phase.

A plausible explanation is that, unlike ReAct's end-to-end generative selection, LLM-BERT delegates action selection to specialized masked language models (type classifier and parameter filler). Small shifts in the presented valid-action candidates and surface-form identifiers can change which parameters these modules consider likely, sometimes pushing the pipeline into repetitive discovery choices (as in $T_{seen}$) and sometimes enabling a cleaner progression to goal actions (as in $T_{unseen}$). When the parameter filler produces suboptimal or repetitive choices, trajectories can still stall and accrue step costs, which helps explain why moderate win rates can coincide with weaker mean returns.

\subsection{Generalization and adaptation agents}
\label{sec:behav_adapt}
Finally, we analyze agents that explicitly target transfer under IP reassignment, either by abstraction (conceptual) or by meta-learning (MAML/Reptile).

\subsubsection{Conceptual Q-learning}
\label{sec:behav_conceptual}
\begin{figure}[!t]
\centering
\begin{subfigure}[t]{\columnwidth}
\centering
\includegraphics[width=\columnwidth,height=0.18\textheight,keepaspectratio]{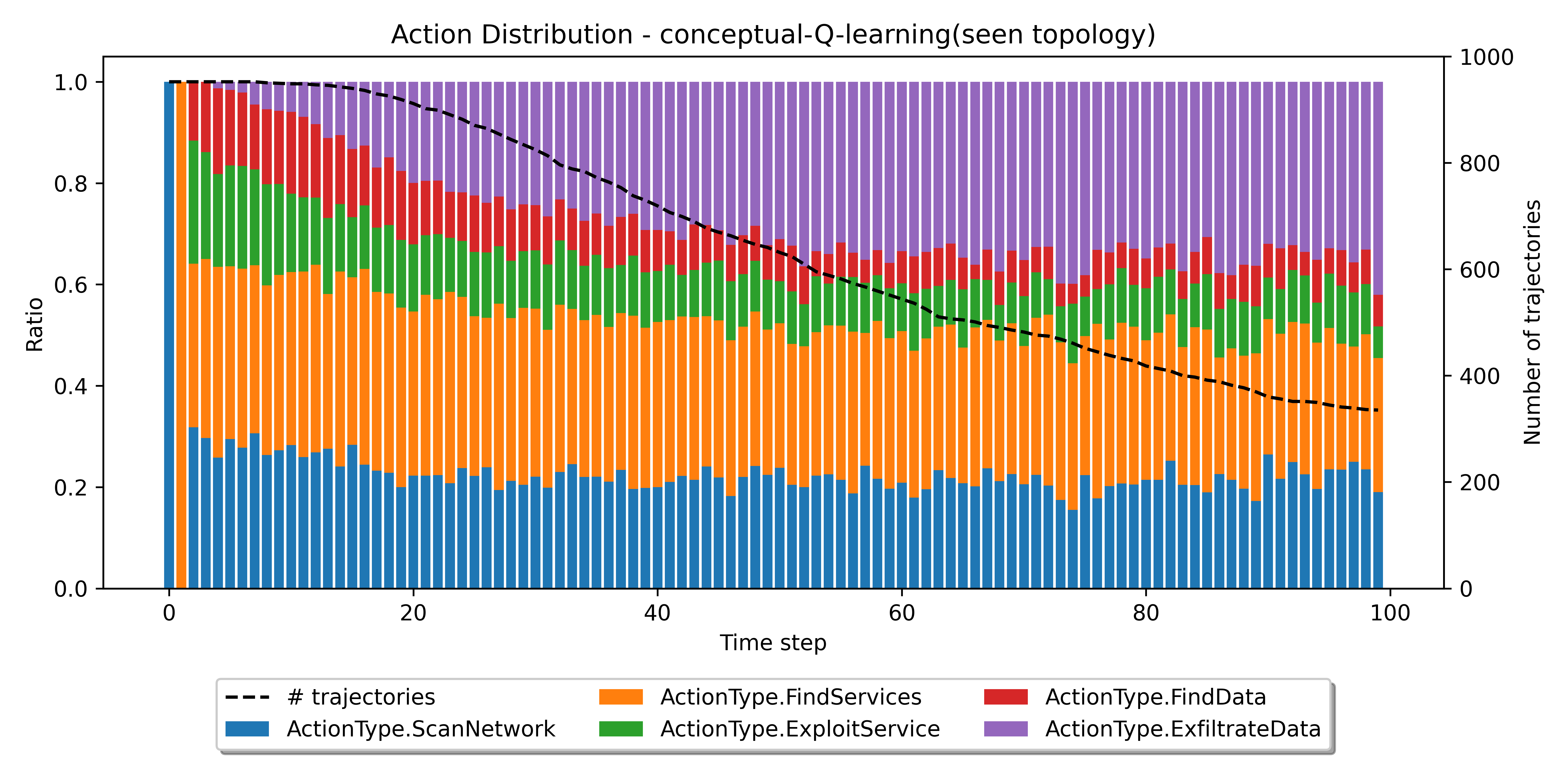}
\caption{Known topology ($T_{seen}$)}
\label{fig:concept_seen}
\end{subfigure}
\vspace{0.5ex}
\begin{subfigure}[t]{\columnwidth}
\centering
\includegraphics[width=\columnwidth,height=0.18\textheight,keepaspectratio]{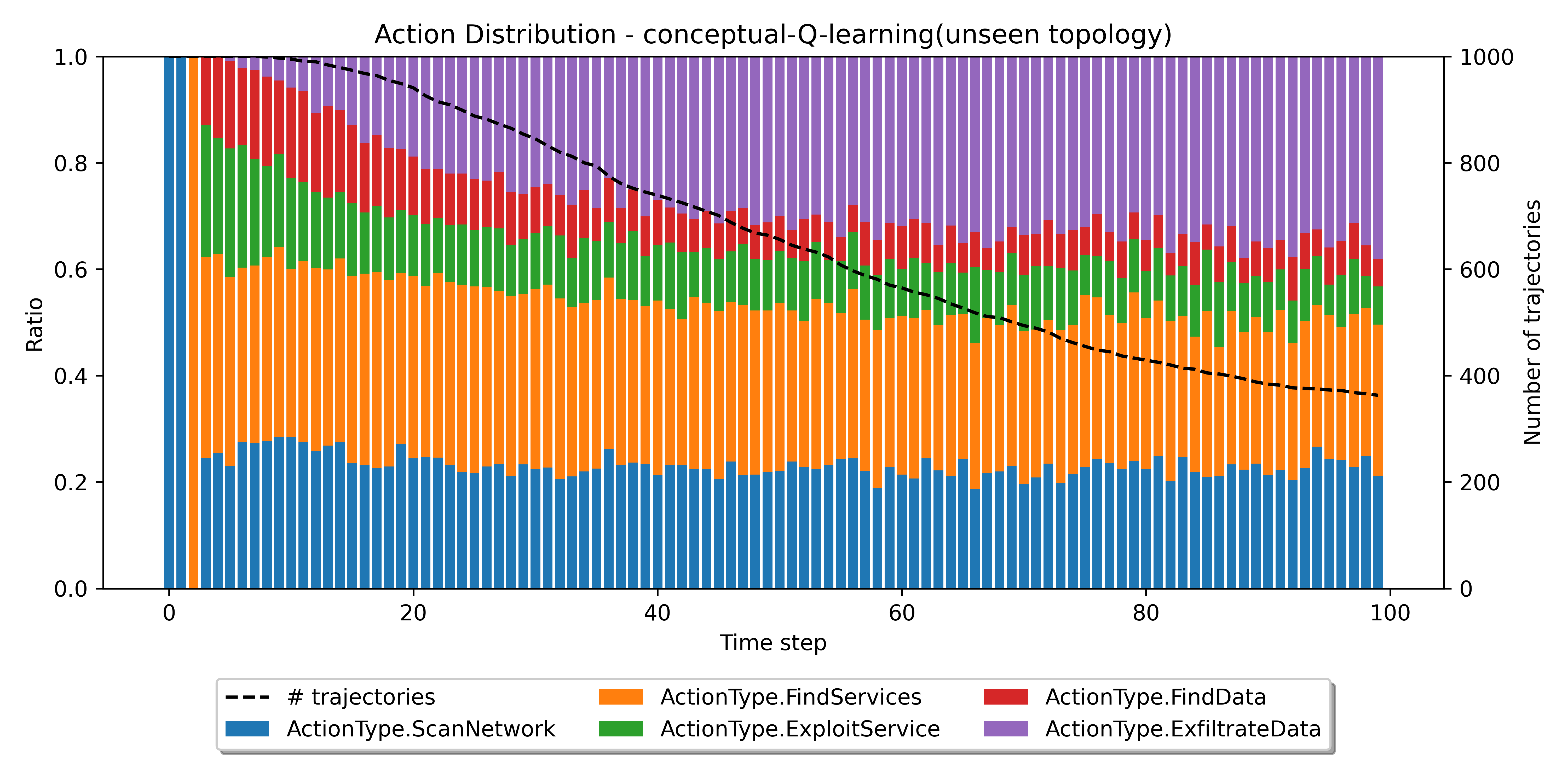}
\caption{Unseen topology ($T_{unseen}$)}
\label{fig:concept_unseen}
\end{subfigure}
\caption{Conceptual Q-learning: action-type distribution over time on seen vs. unseen topology.}
\label{fig:behav_conceptual_pair}
\end{figure}

The conceptual agent exhibits the most consistent seen-to-unseen signature among learning-based approaches (Figures~\ref{fig:concept_seen}--\ref{fig:concept_unseen}). Because its state and action representations abstract away concrete IPs, the same high-level phase structure (discovery $\rightarrow$ exploitation $\rightarrow$ exfiltration) remains visible on $T_{unseen}$. This supports the interpretation that conceptual abstraction preserves the semantics needed for transfer.

The main difference is efficiency: compared to ReAct, the conceptual agent tends to keep a non-trivial mass on reconnaissance actions deeper into the episode, which matches its longer average episode length even when it succeeds.

\subsubsection{MAML}
\label{sec:behav_maml}
\begin{figure}[!t]
\centering
\begin{subfigure}[t]{\columnwidth}
\centering
\includegraphics[width=\columnwidth,height=0.18\textheight,keepaspectratio]{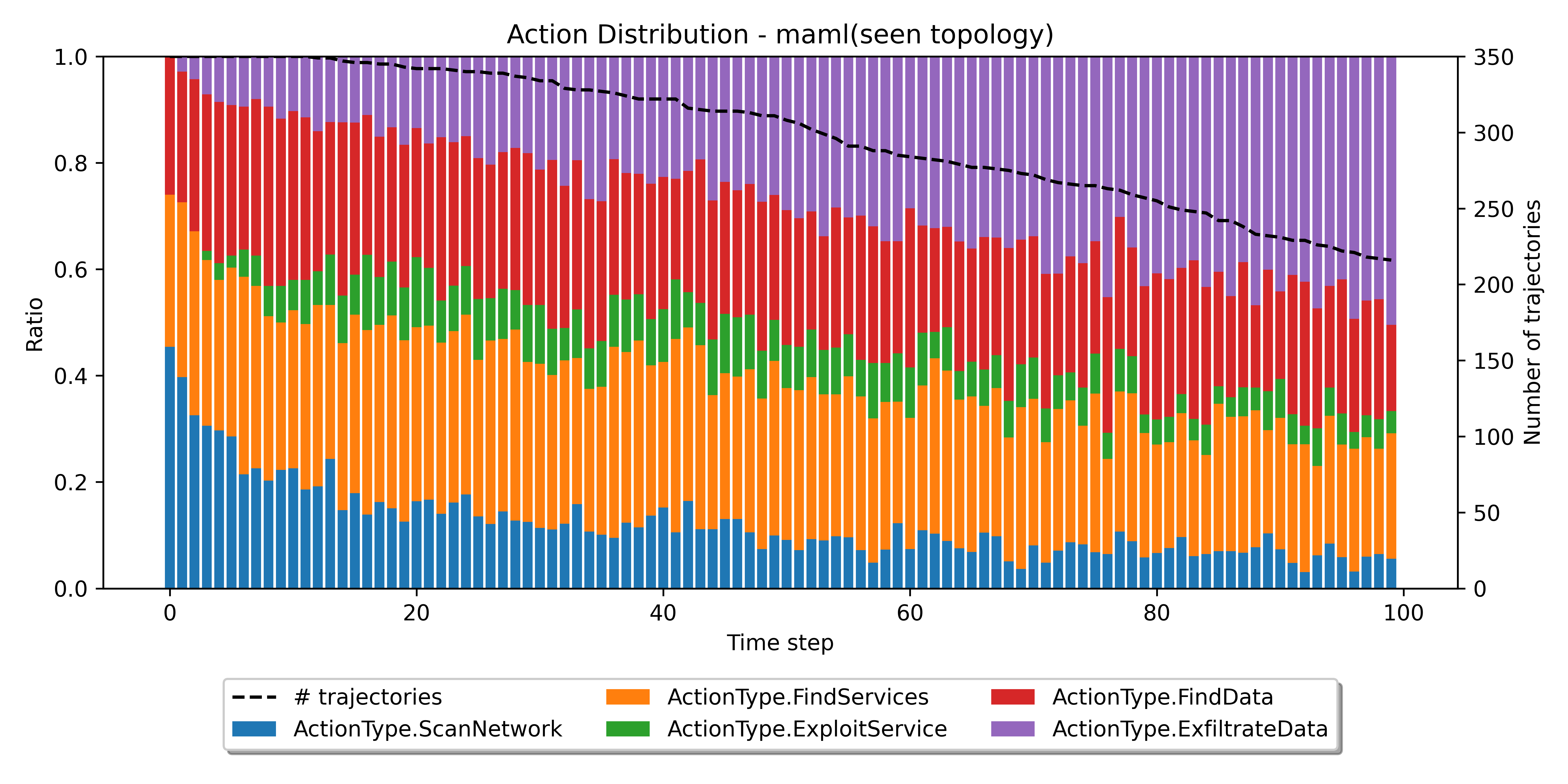}
\caption{Known topology ($T_{seen}$)}
\label{fig:maml_seen}
\end{subfigure}
\vspace{0.5ex}
\begin{subfigure}[t]{\columnwidth}
\centering
\includegraphics[width=\columnwidth,height=0.18\textheight,keepaspectratio]{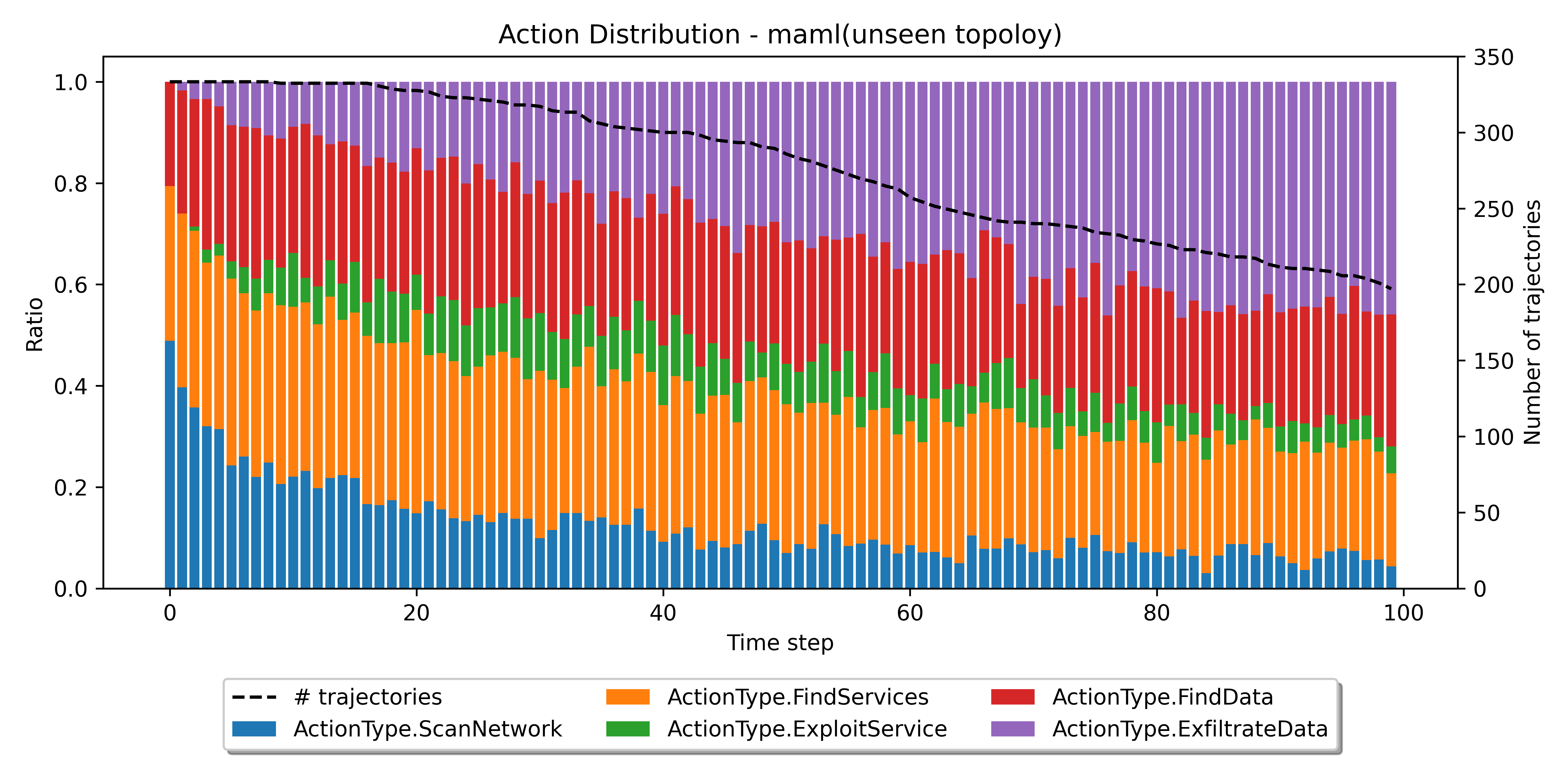}
\caption{Unseen topology ($T_{unseen}$)}
\label{fig:maml_unseen}
\end{subfigure}
\caption{MAML: action-type distribution over time on seen vs. unseen topology (after test-time adaptation).}
\label{fig:behav_maml_pair}
\end{figure}

MAML exhibits only partial preservation of phase structure under reassignment (Figures~\ref{fig:maml_seen}--\ref{fig:maml_unseen}). While there is some increase in downstream actions relative to purely failing value learners, the $T_{unseen}$ signature remains more reconnaissance-heavy and trajectories often persist to later steps. This suggests that the learned initialization plus limited adaptation improves early decision making but does not consistently recover a high-quality long-horizon plan in the unseen task.

\subsubsection{Reptile}
\label{sec:behav_reptile}
\begin{figure}[!t]
\centering
\begin{subfigure}[t]{\columnwidth}
\centering
\includegraphics[width=\columnwidth,height=0.18\textheight,keepaspectratio]{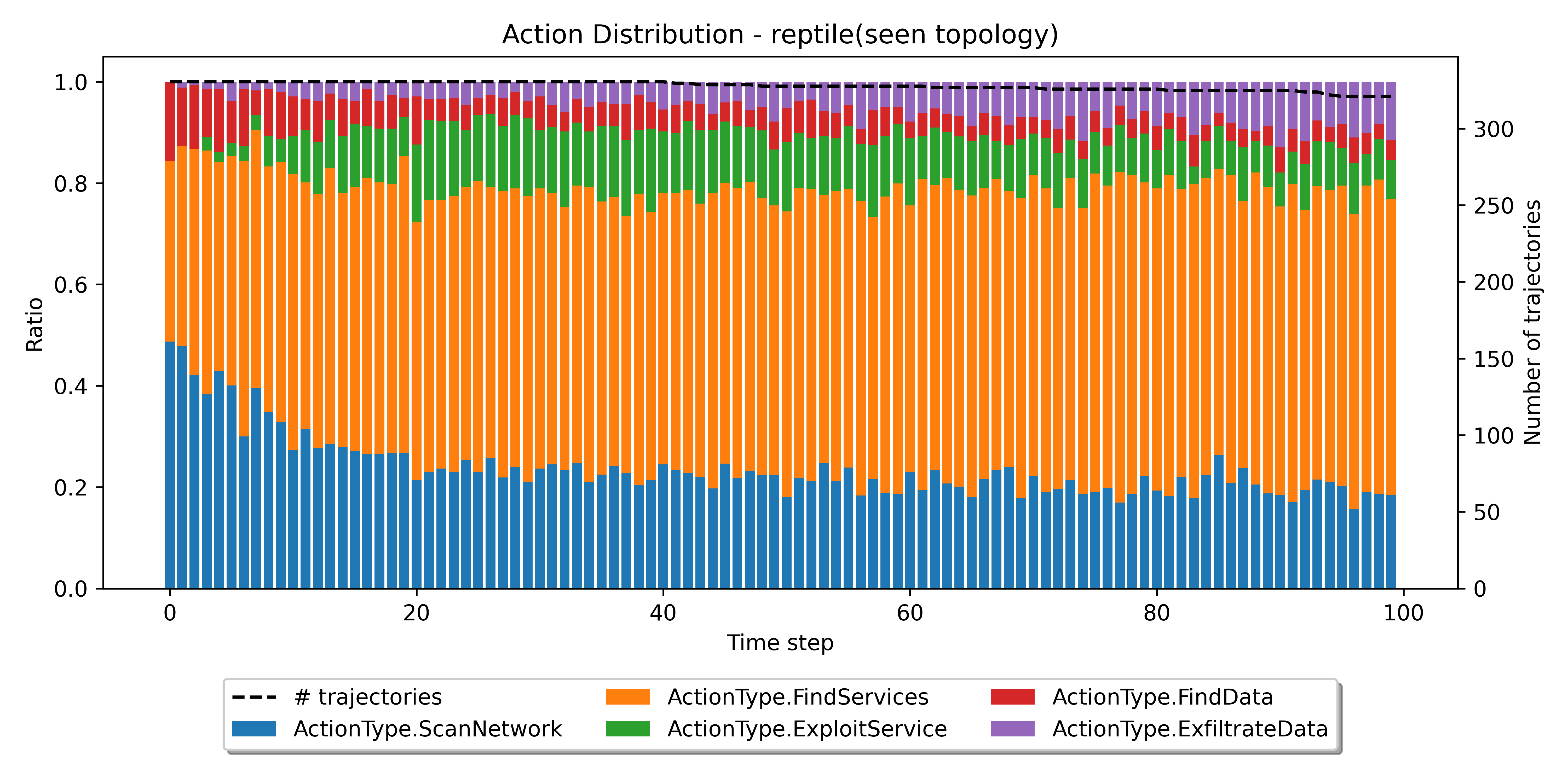}
\caption{Known topology ($T_{seen}$)}
\label{fig:reptile_seen}
\end{subfigure}
\vspace{0.5ex}
\begin{subfigure}[t]{\columnwidth}
\centering
\includegraphics[width=\columnwidth,height=0.18\textheight,keepaspectratio]{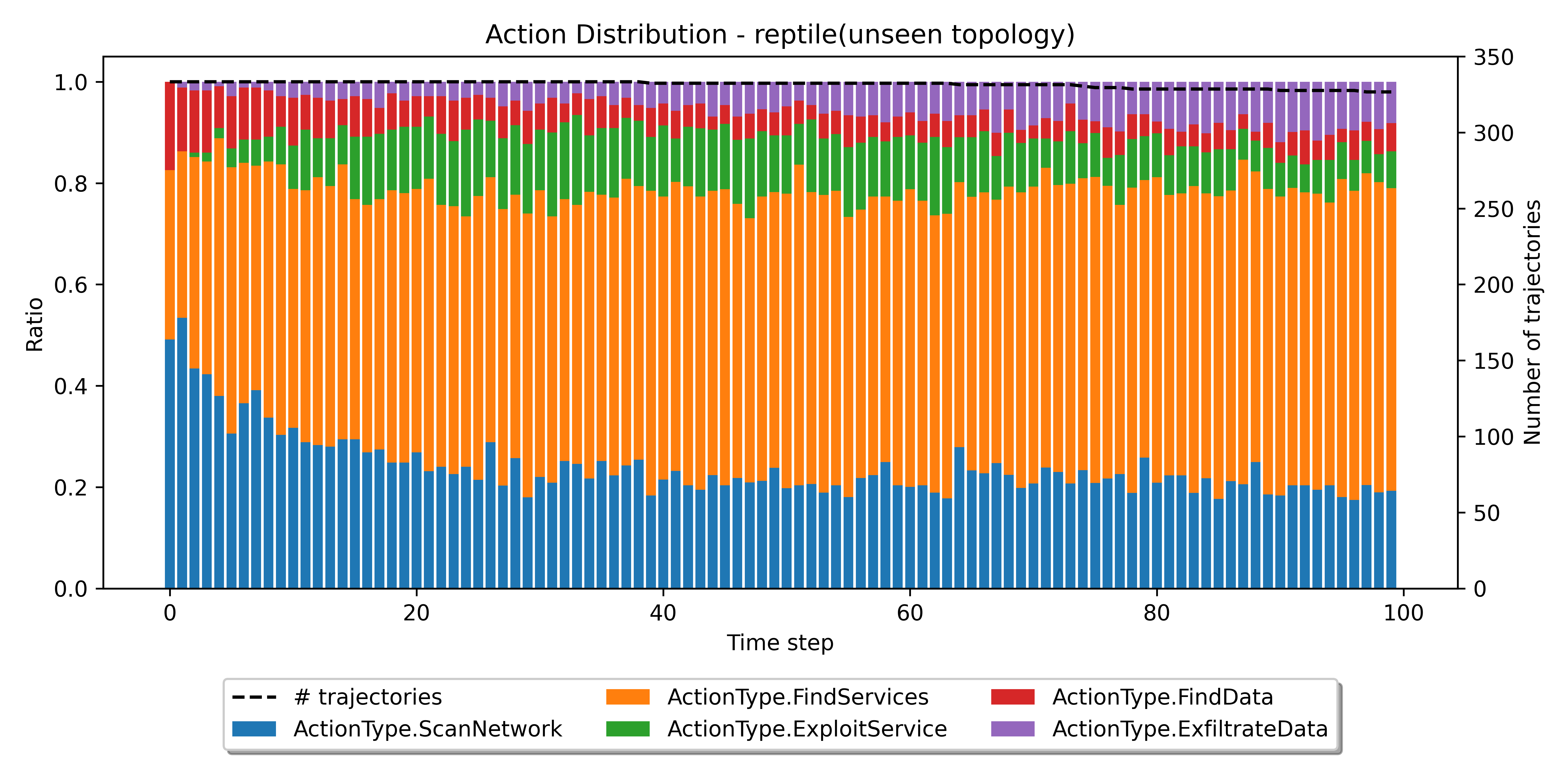}
\caption{Unseen topology ($T_{unseen}$)}
\label{fig:reptile_unseen}
\end{subfigure}
\caption{Reptile: action-type distribution over time on seen vs. unseen topology (after test-time adaptation).}
\label{fig:behav_reptile_pair}
\end{figure}

Reptile most closely resembles the failing value-learning baselines. Its $T_{unseen}$ signature shows limited transition into later-stage actions and sustained reachability, indicating many episodes run close to the timeout. This suggests that the first-order meta-learning updates did not produce an initialization that can be reliably adapted into an effective policy under this task distribution and feature representation.

\subsection{Summary of behavioral comparisons}
Across agent families, the behavioral signatures reveal two distinct failure modes under IP reassignment: (i) \emph{representation collapse}, where the policy remains trapped in early reconnaissance without transitioning to exploitation/exfiltration (DQN/DDQN/Reptile), and (ii) \emph{interface-induced stalling}, where the agent has an appropriate high-level phase structure but can still run long episodes due to loops or invalid actions (ReAct and, to a greater extent, LLM-BERT). Conceptual abstraction most consistently preserves phase structure on $T_{unseen}$, though at the cost of longer trajectories.

\section{Limitations and Future Work}
This section summarizes the main limitations of our study design and experimental scope, and outlines concrete directions to strengthen the generality and interpretability of the findings.

\noindent\textbf{Comparability caveat.} The agents we compare are not compute- or privilege-matched, and should be interpreted as end-to-end designs rather than as tightly controlled algorithmic ablations. As summarized in Table~\ref{tab:comparability}, some methods rely on substantial pretrained priors (LLMs and embedding encoders), some incorporate explicit engineering (abstraction layers, action filters, reward shaping), and some are granted additional interaction at test time via adaptation (meta-learning). We therefore avoid attributing performance differences to any single component in isolation; instead, we use the comparison to map out a robustness--cost--engineering trade-off landscape under a shared environment and evaluation protocol.

Our experiments isolate a single, controlled form of generalization: host and subnet IP reassignment. This makes it possible to evaluate dependence on identifiers, but it does not cover other sources of change, such as network structure and scale, or adaptive defenders. In addition, all results are obtained in NetSecGame and on variants of the same underlying enterprise scenario, so further validation on additional environments and scenario families is needed to assess whether the observed failure modes and successes transfer beyond this benchmark.

The evaluation also includes only a limited set of unseen IP-range variants. Testing on a larger set of held-out reassignments and reporting variability across them would strengthen conclusions about robustness. On the method side, we do not include several promising architecture families designed for permutation invariance and relational reasoning, such as~\cite{Janisch2023a}, which would provide a stronger test of address-invariant representations. 

Finally, comparisons across agent families are complicated by differences in training regimes and, for meta-learning, by test-time adaptation privileges; future work should better control compute budgets and hyperparameter search effort. For LLM-based agents specifically, performance can be sensitive to prompting, decoding settings, and invalid-action handling, so standardizing these choices and reporting ablations would improve reproducibility. Moreover, LLM evaluations should explicitly report inference-time compute (tokens, latency), monetary cost, and any use of closed/proprietary models to make the robustness--cost trade-off transparent.

\section{Conclusions}

We studied whether attacker policies can generalize under a minimal but practically relevant distribution shift: \emph{unseen identifier reassignment} in an otherwise fixed enterprise scenario in NetSecGame. By training on five identifier-variant configurations and evaluating on a sixth, held-out reassignment, we isolated the effect of identifier changes on long-horizon attack planning.

Our results support \textbf{H1} (identifier-dependent policies fail). Traditional value-based agents that bind their internal representations to concrete identifiers exhibit severe seen-to-unseen degradation, including complete failure on $T_{unseen}$ for the embedding-based DDQN baseline. The behavioral-signature analyses further show that these agents often collapse into prolonged reconnaissance without transitioning to exploitation and exfiltration phases, indicating a representation-level mismatch rather than a lack of action availability.

We also find evidence for \textbf{H2} (abstraction and adaptation reduce the drop), but with clear trade-offs. Conceptual Q-learning, which explicitly removes address information through role-based abstraction, transfers substantially better and preserves the multi-phase structure of successful attacks on the unseen topology, albeit with longer trajectories and higher training cost. Meta-learning (MAML/Reptile) provides at best partial recovery in this setting: adaptation improves over purely failing baselines but does not consistently reconstruct high-quality long-horizon behavior under reassignment.

Under the assumptions of our evaluation protocol and agent privileges (Table~\ref{tab:comparability}), prompt-driven \emph{pretrained} LLM agents show the strongest overall robustness to identifier shift among the methods we evaluated. ReAct achieves high win rates and maintains coherent phase structure across reassignments, suggesting that on-the-fly reasoning over discovered evidence can compensate for identifier permutation even without any task-specific policy training. This gain is not ``free,'' however: LLM decision making remains largely a \emph{black box}, often requires substantially more compute per decision, and can incur direct monetary cost (especially with proprietary APIs). In our setting, LLM agents also still exhibit practical failure modes (e.g., repetitive or invalid-action loops) that waste steps, time, and budget. These observations point to a clear engineering agenda: better action grounding and state tracking, explicit repetition/loop control, and cost-aware policies that decide when LLM calls are worth it.

Overall, this study demonstrates that even ``cosmetic'' identifier renaming can break cyber attack policies that implicitly memorize identifiers, whereas pretrained LLM-based reasoning can often recover effective behavior under the same shift. By separating aggregate metrics from behavioral signatures, we provide a diagnostic lens for distinguishing \emph{representation collapse} from \emph{execution-time stalling}. This controlled evaluation setting can help future work design and test identifier-invariant representations, action interfaces, and adaptation mechanisms---and benchmark them against LLM baselines under realistic constraints on compute, transparency, and operational cost.

% references
\bibliographystyle{plain} 
\bibliography{references}
\end{document}